\documentclass[twocolumn]{aastex61}

\usepackage{verbatim, graphicx,  ifthen, pbox}
\newcommand{\be}{\begin{equation}}
\newcommand{\ee}{\end{equation}}
\newcommand{\ThingOne}{2016 QV$_{89}$}
\newcommand{\ThingTwo}{2016 QU$_{89}$}
\usepackage{xcolor}
\usepackage{amsmath}
\usepackage{graphicx}

\usepackage{verbatim}
\usepackage{booktabs}

\shorttitle{Three Similar ETNOs}
\shortauthors{Khain et al.}

\begin{document}

\title{Dynamical Analysis of Three Distant Trans-Neptunian Objects with Similar Orbits}

\correspondingauthor{Tali Khain}
\email{talikh@umich.edu}

\author[0000-0001-7721-6457]{T.~Khain}
\affiliation{Department of Physics, University of Michigan, Ann Arbor, MI 48109, USA}
\author[0000-0002-7733-4522]{J.~C.~Becker}
\altaffiliation{NSF Graduate Research Fellow}
\affiliation{Department of Astronomy, University of Michigan, Ann Arbor, MI 48109, USA}

\author[0000-0002-8167-1767]{F.~C.~Adams}
\affiliation{Department of Physics, University of Michigan, Ann Arbor, MI 48109, USA}
\affiliation{Department of Astronomy, University of Michigan, Ann Arbor, MI 48109, USA}
\author[0000-0001-6942-2736]{D.~W.~Gerdes}
\affiliation{Department of Physics, University of Michigan, Ann Arbor, MI 48109, USA}
\affiliation{Department of Astronomy, University of Michigan, Ann Arbor, MI 48109, USA}
\author[0000-0002-6126-8487]{S.~J.~Hamilton}
\altaffiliation{NSF Graduate Research Fellow}
\affiliation{Department of Physics, University of Michigan, Ann Arbor, MI 48109, USA}
\author[0000-0002-8906-2835]{K.~Franson}
\affiliation{Department of Physics, University of Michigan, Ann Arbor, MI 48109, USA}
\author[0000-0002-8906-2835]{L.~Zullo}
\affiliation{Department of Physics, University of Michigan, Ann Arbor, MI 48109, USA}

\author[0000-0003-2764-7093]{M.~Sako}
\affiliation{Department of Physics and Astronomy, University of Pennsylvania, Philadelphia, PA 19104, USA}
\author{K. Napier}
\affiliation{Department of Physics, University of Michigan, Ann Arbor, MI 48109, USA}
\author[0000-0001-7737-6784]{Hsing~Wen~Lin}
\affiliation{Department of Physics, University of Michigan, Ann Arbor, MI 48109, USA}
\author[0000-0002-2486-1118]{L. Markwardt} 
\altaffiliation{NSF Graduate Research Fellow}
\affiliation{Department of Astronomy, University of Michigan, Ann Arbor, MI 48109, USA}
\author[0000-0003-0743-9422]{P.~Bernardinelli}
\affiliation{Department of Physics and Astronomy, University of Pennsylvania, Philadelphia, PA 19104, USA}

\author{T.~M.~C.~Abbott}
\affiliation{Cerro Tololo Inter-American Observatory, National Optical Astronomy Observatory, Casilla 603, La Serena, Chile}
\author{F.~B.~Abdalla}
\affiliation{Department of Physics and Electronics, Rhodes University, PO Box 94, Grahamstown, 6140, South Africa}
\affiliation{Department of Physics \& Astronomy, University College London, Gower Street, London, WC1E 6BT, UK}
\author{J.~Annis}
\affiliation{Fermi National Accelerator Laboratory, P. O. Box 500, Batavia, IL 60510, USA}
\author{S.~Avila}
\affiliation{Institute of Cosmology \& Gravitation, University of Portsmouth, Portsmouth, PO1 3FX, UK}
\author{E.~Bertin}
\affiliation{Sorbonne Universit\'es, UPMC Univ Paris 06, UMR 7095, Institut d'Astrophysique de Paris, F-75014, Paris, France}
\affiliation{CNRS, UMR 7095, Institut d'Astrophysique de Paris, F-75014, Paris, France}
\author{D.~Brooks}
\affiliation{Department of Physics \& Astronomy, University College London, Gower Street, London, WC1E 6BT, UK}
\author{A.~Carnero~Rosell}
\affiliation{Laborat\'orio Interinstitucional de e-Astronomia - LIneA, Rua Gal. Jos\'e Cristino 77, Rio de Janeiro, RJ - 20921-400, Brazil}
\affiliation{Observat\'orio Nacional, Rua Gal. Jos\'e Cristino 77, Rio de Janeiro, RJ - 20921-400, Brazil}
\author{M.~Carrasco~Kind}
\affiliation{Department of Astronomy, University of Illinois at Urbana-Champaign, 1002 W. Green Street, Urbana, IL 61801, USA}
\affiliation{National Center for Supercomputing Applications, 1205 West Clark St., Urbana, IL 61801, USA}
\author{J.~Carretero}
\affiliation{Institut de F\'{\i}sica d'Altes Energies (IFAE), The Barcelona Institute of Science and Technology, Campus UAB, 08193 Bellaterra (Barcelona) Spain}
\author{C.~E.~Cunha}
\affiliation{Kavli Institute for Particle Astrophysics \& Cosmology, P. O. Box 2450, Stanford University, Stanford, CA 94305, USA}
\author{L.~N.~da Costa}
\affiliation{Observat\'orio Nacional, Rua Gal. Jos\'e Cristino 77, Rio de Janeiro, RJ - 20921-400, Brazil}
\affiliation{Laborat\'orio Interinstitucional de e-Astronomia - LIneA, Rua Gal. Jos\'e Cristino 77, Rio de Janeiro, RJ - 20921-400, Brazil}
\author{C.~Davis}
\affiliation{Kavli Institute for Particle Astrophysics \& Cosmology, P. O. Box 2450, Stanford University, Stanford, CA 94305, USA}
\author{J.~De~Vicente}
\affiliation{Centro de Investigaciones Energ\'eticas, Medioambientales y Tecnol\'ogicas (CIEMAT), Madrid, Spain}
\author{S.~Desai}
\affiliation{Department of Physics, IIT Hyderabad, Kandi, Telangana 502285, India}
\author{H.~T.~Diehl}
\affiliation{Fermi National Accelerator Laboratory, P. O. Box 500, Batavia, IL 60510, USA}
\author{P.~Doel}
\affiliation{Department of Physics \& Astronomy, University College London, Gower Street, London, WC1E 6BT, UK}
\author{T.~F.~Eifler}
\affiliation{Department of Astronomy/Steward Observatory, 933 North Cherry Avenue, Tucson, AZ 85721-0065, USA}
\affiliation{Jet Propulsion Laboratory, California Institute of Technology, 4800 Oak Grove Dr., Pasadena, CA 91109, USA}
\author{B.~Flaugher}
\affiliation{Fermi National Accelerator Laboratory, P. O. Box 500, Batavia, IL 60510, USA}
\author{J.~Frieman}
\affiliation{Kavli Institute for Cosmological Physics, University of Chicago, Chicago, IL 60637, USA}
\affiliation{Fermi National Accelerator Laboratory, P. O. Box 500, Batavia, IL 60510, USA}
\author{J.~Garc\'ia-Bellido}
\affiliation{Instituto de Fisica Teorica UAM/CSIC, Universidad Autonoma de Madrid, 28049 Madrid, Spain}
\author{D.~Gruen}
\affiliation{SLAC National Accelerator Laboratory, Menlo Park, CA 94025, USA}
\affiliation{Kavli Institute for Particle Astrophysics \& Cosmology, P. O. Box 2450, Stanford University, Stanford, CA 94305, USA}
\author{R.~A.~Gruendl}
\affiliation{Department of Astronomy, University of Illinois at Urbana-Champaign, 1002 W. Green Street, Urbana, IL 61801, USA}
\affiliation{National Center for Supercomputing Applications, 1205 West Clark St., Urbana, IL 61801, USA}
\author{G.~Gutierrez}
\affiliation{Fermi National Accelerator Laboratory, P. O. Box 500, Batavia, IL 60510, USA}
\author{W.~G.~Hartley}
\affiliation{Department of Physics, ETH Zurich, Wolfgang-Pauli-Strasse 16, CH-8093 Zurich, Switzerland}
\affiliation{Department of Physics \& Astronomy, University College London, Gower Street, London, WC1E 6BT, UK}
\author{D.~L.~Hollowood}
\affiliation{Santa Cruz Institute for Particle Physics, Santa Cruz, CA 95064, USA}
\author{K.~Honscheid}
\affiliation{Center for Cosmology and Astro-Particle Physics, The Ohio State University, Columbus, OH 43210, USA}
\affiliation{Department of Physics, The Ohio State University, Columbus, OH 43210, USA}
\author{D.~J.~James}
\affiliation{Harvard-Smithsonian Center for Astrophysics, Cambridge, MA 02138, USA}
\author{E.~Krause}
\affiliation{Department of Astronomy/Steward Observatory, 933 North Cherry Avenue, Tucson, AZ 85721-0065, USA}
\author{K.~Kuehn}
\affiliation{Australian Astronomical Observatory, North Ryde, NSW 2113, Australia}
\author{N.~Kuropatkin}
\affiliation{Fermi National Accelerator Laboratory, P. O. Box 500, Batavia, IL 60510, USA}
\author{O.~Lahav}
\affiliation{Department of Physics \& Astronomy, University College London, Gower Street, London, WC1E 6BT, UK}
\author{M.~A.~G.~Maia}
\affiliation{Laborat\'orio Interinstitucional de e-Astronomia - LIneA, Rua Gal. Jos\'e Cristino 77, Rio de Janeiro, RJ - 20921-400, Brazil}
\affiliation{Observat\'orio Nacional, Rua Gal. Jos\'e Cristino 77, Rio de Janeiro, RJ - 20921-400, Brazil}
\author{F.~Menanteau}
\affiliation{Department of Astronomy, University of Illinois at Urbana-Champaign, 1002 W. Green Street, Urbana, IL 61801, USA}
\affiliation{National Center for Supercomputing Applications, 1205 West Clark St., Urbana, IL 61801, USA}
\author{R.~Miquel}
\affiliation{Instituci\'o Catalana de Recerca i Estudis Avan\c{c}ats, E-08010 Barcelona, Spain}
\affiliation{Institut de F\'{\i}sica d'Altes Energies (IFAE), The Barcelona Institute of Science and Technology, Campus UAB, 08193 Bellaterra (Barcelona) Spain}
\author{B.~Nord}
\affiliation{Fermi National Accelerator Laboratory, P. O. Box 500, Batavia, IL 60510, USA}
\author{R.~L.~C.~Ogando}
\affiliation{Observat\'orio Nacional, Rua Gal. Jos\'e Cristino 77, Rio de Janeiro, RJ - 20921-400, Brazil}
\affiliation{Laborat\'orio Interinstitucional de e-Astronomia - LIneA, Rua Gal. Jos\'e Cristino 77, Rio de Janeiro, RJ - 20921-400, Brazil}
\author{A.~A.~Plazas}
\affiliation{Jet Propulsion Laboratory, California Institute of Technology, 4800 Oak Grove Dr., Pasadena, CA 91109, USA}
\author{A.~K.~Romer}
\affiliation{Department of Physics and Astronomy, Pevensey Building, University of Sussex, Brighton, BN1 9QH, UK}
\author{E.~Sanchez}
\affiliation{Centro de Investigaciones Energ\'eticas, Medioambientales y Tecnol\'ogicas (CIEMAT), Madrid, Spain}
\author{V.~Scarpine}
\affiliation{Fermi National Accelerator Laboratory, P. O. Box 500, Batavia, IL 60510, USA}
\author{R.~Schindler}
\affiliation{SLAC National Accelerator Laboratory, Menlo Park, CA 94025, USA}
\author{M.~Schubnell}
\affiliation{Department of Physics, University of Michigan, Ann Arbor, MI 48109, USA}
\author{I.~Sevilla-Noarbe}
\affiliation{Centro de Investigaciones Energ\'eticas, Medioambientales y Tecnol\'ogicas (CIEMAT), Madrid, Spain}
\author{M.~Smith}
\affiliation{School of Physics and Astronomy, University of Southampton,  Southampton, SO17 1BJ, UK}
\author{M.~Soares-Santos}
\affiliation{Brandeis University, Physics Department, 415 South Street, Waltham MA 02453}
\author{F.~Sobreira}
\affiliation{Instituto de F\'isica Gleb Wataghin, Universidade Estadual de Campinas, 13083-859, Campinas, SP, Brazil}
\affiliation{Laborat\'orio Interinstitucional de e-Astronomia - LIneA, Rua Gal. Jos\'e Cristino 77, Rio de Janeiro, RJ - 20921-400, Brazil}
\author{E.~Suchyta}
\affiliation{Computer Science and Mathematics Division, Oak Ridge National Laboratory, Oak Ridge, TN 37831}
\author{M.~E.~C.~Swanson}
\affiliation{National Center for Supercomputing Applications, 1205 West Clark St., Urbana, IL 61801, USA}
\author{G.~Tarle}
\affiliation{Department of Physics, University of Michigan, Ann Arbor, MI 48109, USA}
\author{V.~Vikram}
\affiliation{Argonne National Laboratory, 9700 South Cass Avenue, Lemont, IL 60439, USA}
\author{A.~R.~Walker}
\affiliation{Cerro Tololo Inter-American Observatory, National Optical Astronomy Observatory, Casilla 603, La Serena, Chile}
\author{W.~Wester}
\affiliation{Fermi National Accelerator Laboratory, P. O. Box 500, Batavia, IL 60510, USA}
\author{Y.~Zhang}
\affiliation{Fermi National Accelerator Laboratory, P. O. Box 500, Batavia, IL 60510, USA}

\collaboration{(DES Collaboration)}

\begin{abstract}

This paper reports the discovery and orbital characterization of two extreme trans-Neptunian objects (ETNOs), \ThingOne\ and \ThingTwo, which have orbits that appear similar to that of a previously known object, 2013 UH$_{15}$. All three ETNOs have semi-major axes $a\approx 172$~AU and eccentricities $e\approx0.77$. 
The angular elements $(i,\omega,\Omega)$ vary by 6, 15, and 49 deg, respectively between the three objects. 
The two new objects add to the small number of TNOs currently known to have semi-major axes between 150 and 250 AU, and serve as an interesting dynamical laboratory to study the outer realm of our Solar System. 
Using a large ensemble of numerical integrations, we find that the orbits are expected to reside in close proximity in the $(a,e)$ phase plane for roughly 100 Myr before diffusing to more separated values. 
We find that an explanation for the orbital configuration of the bodies as a collision product is disfavored. 
We then explore other scenarios that could influence their orbits. 
With aphelion distances over 300 AU, the orbits of these ETNOs extend far beyond the classical Kuiper Belt, and an order of magnitude beyond Neptune. As a result, their orbital dynamics can be affected by the proposed new Solar System member, referred to as Planet Nine in this work. With perihelion distances of 35 -- 40 AU, these orbits are also influenced by resonant interactions with Neptune. A full assessment of any possible, new Solar System planets must thus take into account this emerging class of TNOs. 

\end{abstract}

\section{Introduction} 

In addition to its major planets, the Solar System contains a vast number of small, rocky bodies with a variety of orbital elements. The orbits of these minor bodies provide an important record of the history of our Solar System. In particular, the history of impacts and binary dissociations, which leave minimal observable traces, can be discerned through dynamical techniques in the asteroid belt \citep[e.g.,][]{2002Natur.417..720N}, among Jovian satellites \citep[e.g.,][]{2003AJ....126..398N}, and in the Kuiper belt \citep[e.g., the case of Haumea;][]{2007Natur.446..294B}.

This paper reports the discovery and dynamical characterization of two new extreme trans-Neptunian objects, and a third previously known body, all of which exhibit similar orbital elements. Unfortunately, the discovery and characterization of objects with common histories becomes more difficult in the outer Solar System (beyond Neptune), where the surface density of known objects is much lower than in the asteroid belt, and objects are often observable only near perihelion. Although an initial velocity dispersion of a few hundred meters per second will result in orbits that disperse quickly, the semi-major axes, eccentricities, and inclinations of such objects are expected to remain differentiable from the background \citep[see Figures 2 and 3 of][]{2011ApJ...733...40M}. As a result, groups of objects with similar orbital elements require additional study to discern whether or not they actually have a common origin. Haumea is the best example of a Kuiper Belt collisional family \citep{2009AJ....137.4766R,2009ApJ...700.1242S,2012Icar..221..106V}. Since its initial discovery, more family members (which have orbits similar to each other) have been found and confirmed. 

Among the population of trans-Neptunian objects (TNOs), unidentified families certainly exist. 
Recent progress towards identifying such associated objects in the outer Solar System comes from the work of \citet{2018MNRAS.474..838D}, who performed a correlation analysis on outer Solar System objects and found a number of potentially associated objects. One leading candidate for an associated pair of ETNOs beyond 150 AU is the case of 2004 VN$_{112}$ and 2013 RF$_{98}$ \citep{2017MNRAS.467L..66D}, which have been proposed to have come from a binary dissociation event \citep{2017Ap&SS.362..198D}. 
The identification of additional associated objects in the Solar System beyond Neptune will enable a better understanding of binary dissociation mechanisms at all orbital locations. 


In recent years, dedicated  \citep{2004AJ....128.1364B,2017AJ....153..236P} and serendipitous surveys \citep{2004ApJ...610.1199G, 2015MNRAS.446..932L, 2017ApJ...839L..15G} have allowed for the discovery of many new objects with more distant orbits than was previously thought possible \citep{2001ApJ...549L.241A}, allowing for the identification of new trends.
\citet{st14} observed an alignment in argument of perihelion for the most distant TNOs, and \citet{BB16} subsequently pointed out an additional alignment in longitude of perihelion. This clustering was used as evidence for the Planet Nine hypothesis. Since then, an additional group of eight extreme TNOs in this class (with $a>250$ AU) has been found. 
Less distant TNOs, with semi-major axes between 150 AU to 250 AU, can also provide insight towards the Planet Nine hypothesis, as it remains unclear exactly where the demarcation between the TNOs influenced mainly by Neptune and those influenced primarily by Planet Nine should lie.
As a result, the identification of new ETNOs with semi-major axes in the range $a$ = 150--250 AU not only expands our census of the outer Solar System, but provides further constraints on the Planet Nine hypothesis. 

In this paper, we use data from the Dark Energy Survey (DES) to discover two new objects in the outer Kuiper Belt with large semi-major axes. Motivated by the importance of identifying potential family candidates in the outer Solar System, we examine the apparent similarities between their orbits with each other and with another previously-discovered Kuiper Belt Object (KBO). With semi-major axes $a\sim170$ AU and aphelion distances $\sim300$ AU, these new objects add to the growing inventory of distant objects in the Solar System. Motivated by the uniqueness of the large orbital distances (as these objects are the tenth and eleventh known objects to have semi-major axis between 150 and 250 AU), we also evaluate their orbital dynamics. 

Section \ref{sec:discovery} describes the observational results, including the methods used in DES and the analysis that specifies the orbital elements of the newly discovered bodies. The dynamics of these objects is studied in Section \ref{sec:dynamics}, which considers whether the apparent similarity in their orbits is potentially due to random chance. In Section \ref{sec:resonance}, we investigate the role of mean-motion resonances in the dynamics of these objects, considering both possible resonances with Neptune and with the proposed Planet Nine. The paper ends in Section \ref{sec:conclude} with a summary and a look to the future.

\section{Discovery}
\label{sec:discovery} 

\begin{figure*}[t!]
\epsscale{1}
  \begin{center}
      \leavevmode
\includegraphics[width=176mm]{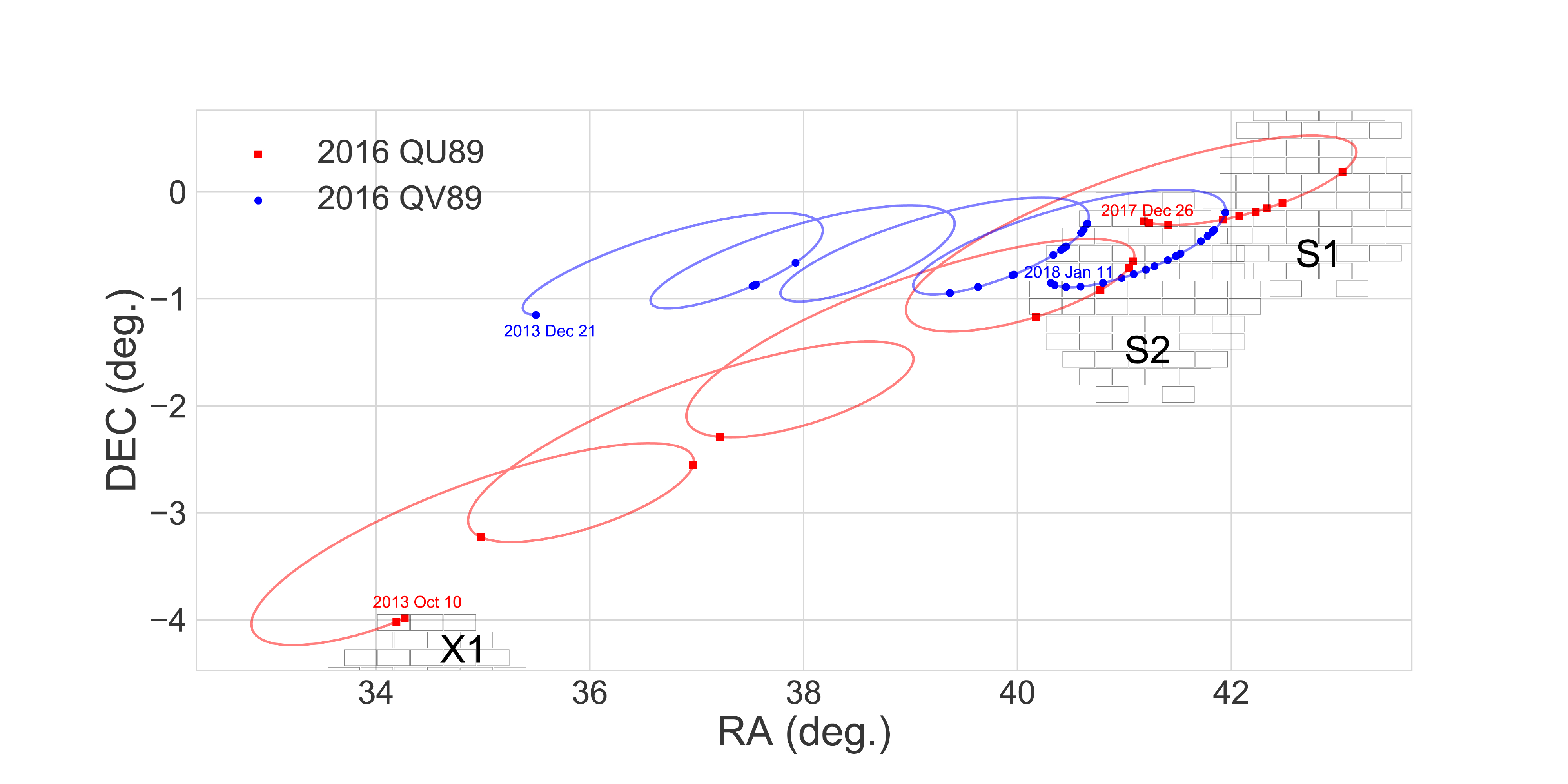}
\caption{The path of \ThingOne\ and \ThingTwo\ over the course of the entire (to-date) DES survey. Nights on which the objects were detected by DES are shown as dots along each trajectory. The objects were initially identified in the S1, S2, and X1 supernova fields, which are indicated by outlines of the DECam focal plane. Additional observations were subsequently added from wide survey exposures outside these fields.}
\label{fig:discovery}
\end{center}
\end{figure*}

The Dark Energy Survey (DES) \citep{des_survey} is an optical survey that observes nearly 5000 square degrees of the southern sky using the 4-meter Blanco telescope at the Cerro Tololo Inter-American Observatory in Chile. DES had a nominal survey allocation of 525 nights over five years from September 2013 through February 2018; a portion of a sixth season is planned for late 2018. 
DES uses the Dark Energy Camera (DECam) \citep{DECam2015}, a prime-focus camera with a 3 square degree field of view and a focal plane consisting of 62 separate $2K \times 4K$ red-sensitive CCDs.
There are two distinct DES survey modes: the wide survey and the supernova survey \citep{2012ApJ...753..152B}. 
Wide field survey fields are observed roughly 2-4 times per 6-month observing campaign in each of the five \emph{grizY} photometric bands, to nominal single-exposure 10$\sigma$ depths of $g = 23.6$, $r = 23.3$, 
$i = 22.8$, $z = 22.1$, and $Y = 20.7$ \citep{DECam-pipeline-2018}, 
with the result being eventual periodic complete coverage of the 5000 square degree survey area in each wavelength band. 
In contrast, the supernova survey comprises ten distinct DECam pointings where repeated observations in the $griz$ bands are made roughly every 6 days. 
The ten supernova fields are small (3 square degrees each) compared to the wide field (which makes up the remainder of the 5000 square degrees of the survey area), but their dense observing cadence and somewhat greater depth makes them well-suited for moving object searches.
So far, TNO discoveries have come from both supernova survey \citep{2016AJ....151...39G} and wide survey \citep{2017ApJ...839L..15G, caju, RedTrojan} DES data. 

\subsection{Detection of \ThingOne\ and \ThingTwo}
\label{sec:discover} 
The original detection of these objects came in supernova survey data from the DES 2016-17 observing campaign (DES Year 4; \citealt{2018SPIE10704E..0DD}).
Data from these fields were processed with the DES difference imaging pipeline \citep{2015AJ....150..172K}. 
A template image was subtracted from each new search image, and statistically significant sources in the subtracted image were identified. Artifacts and non-psf-like sources were rejected using the techniques in \citet{2015AJ....150...82G}, resulting in a relatively clean catalog of single-epoch transients.
We then identified pairs of detections within 20 nights of each other whose angular separation was consistent with what would be expected given the predicted earth-reflex motion of a distant object. 
Once a database of pairs had been constructed, we linked the pairs into chains of observations by testing the goodness of fit of the best-fit orbit to each chain and requiring the reduced chi-squared $\chi^{2}/N < 2$ \citep{2000AJ....120.3323B} to qualify as a detection.  The reduced chi-squared of the fits were 1.75 for \ThingOne, 1.3 for \ThingTwo, and 1.1 for 2013 UH$_{15}$.

After the initial barycentric orbit was determined for both objects, we searched for additional detections in wide-survey and supernova exposures from epochs both before and after the discovery opposition. Indeed, \ThingTwo\ was found to have appeared in a supernova field on two nights in October 2013, less than a month
after the start of the survey. The trajectories of these objects over the full five years of the survey are shown in Figure~\ref{fig:discovery}. The full five-year DES data set allows both orbital arcs over multiple oppositions, resulting in total arc lengths \footnote{The arc length of an object is the number of days between the earliest and latest observations of the object.} of 1481 and 1537 days for \ThingOne\ and \ThingTwo\ respectively. 

\begin{deluxetable*}{clll}
\tablecaption{Orbital elements of the three ETNOs considered in this work.}
\tablewidth{0pt}
\tablehead{
  \colhead{Parameter } &
  \colhead{2013 UH$_{15}$}   &
    \colhead{\ThingOne}   &
  \colhead{\ThingTwo}
}
\startdata
$a$ & 173.6 $\pm$ 1.7 AU & 171.70   $\pm$ 0.05 AU & 171.40  $\pm$ 0.02 AU \\
$e$ & 0.798 $\pm$ 0.002 & 0.76731  $\pm$ 0.00007  & 0.79439  $\pm$ 0.00002 \\
$i$ & 26.081 $\pm$ 0.001 deg & 21.38750  $\pm$ 0.00007 deg & 16.97552  $\pm$ 0.00002 deg \\
$\omega$ & 282.87 $\pm$ 0.06 deg & 281.093  $\pm$ 0.004 deg & 303.337 $\pm$ 0.004 deg \\
$\Omega$ & 176.543 $\pm$ 0.001 deg & 173.2150  $\pm$ 0.0002 deg & 102.8996  $\pm$ 0.0002 deg \\
Epoch & 2456594.5804 JD & 2456647.6445  JD & 2456575.6372 JD \\
Time of Perihelion & 2472269.15 $\pm$ 14.12 JD & 2469915.40  $\pm$ 0.21  JD & 2459260.79  $\pm$ 0.60 JD \\
Perihelion& 35.0 $\pm$  0.7 AU & 39.95  $\pm$ 0.02 AU & 35.249  $\pm$ 0.007 AU \\
Aphelion & 312  $\pm$  3 AU & 303.45  $\pm$ 0.09 AU & 307.63  $\pm$ 0.03 AU\\
Absolute magnitude & 7.7 &  5.9 &  7.95 \\
$\emph{g-r}$ (mag) & - &  0.66 $\pm$ 0.10 &  0.64 $\pm$ 0.09 \\
$\emph{r-z}$ (mag) & - & 0.43 $\pm$ 0.12 & 0.44 $\pm$ 0.17\\
$\emph{i-z}$ (mag) & - &  0.15 $\pm$ 0.18 & 0.17 $\pm$ 0.18 \\
$\emph{r-i}$ (mag) & - & 0.31 $\pm$ 0.11 & 0.23 $\pm$ 0.17 \\
\enddata
\tablecomments{Solution for 2013 UH$_{15}$ was computed 
using data from JPL's SSDG SBDB calculated at epoch JD 2458000.5
as written in Table 1 of \citet{2017MNRAS.471L..61D}. The arc lengths of DES observations were 1481 and 1537 days for \ThingOne\ and \ThingTwo\ respectively. Colors of \ThingOne\ and \ThingTwo\ come from DES's multi-waveband observations. As 2013 UH$_{15}$ has not been observed by DES and the observations used to compute its orbit were all taken in $r$-band, there do not yet exist colors for this object.}
\label{table:1}
\end{deluxetable*}

\begin{figure}[t!]
\epsscale{1}
  \begin{center}
      \leavevmode
\includegraphics[width=85mm]{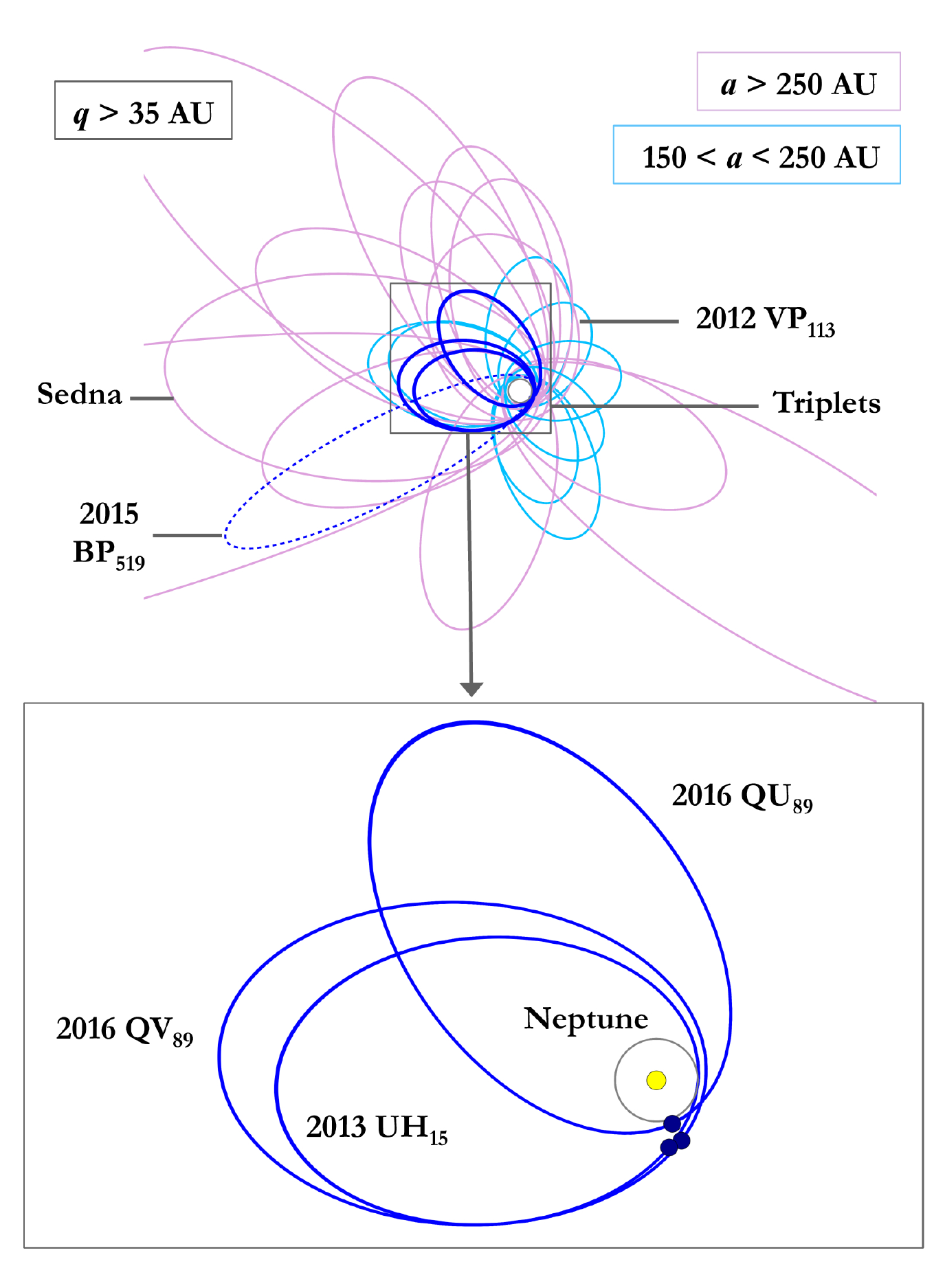}
\caption{The top panel shows the orbits of all currently known ETNOs with $q > 35$ AU and $a > 150$ AU. The objects with $150 < a < 250$ AU are shown in light blue, and objects with $a > 250$ AU are in pink. The three objects we discuss in this work (which we call the triplet objects) are boxed and in blue, and the dashed object is the high-inclination, recently discovered  extreme TNO 2015~BP$_{519}$ \citep{caju}. The bottom panel shows a closer view of the orbital geometries of the triplet objects. The filled-in circles indicate their position on the orbit at discovery (near perihelion).}
\label{fig:orbits}
\end{center}
\end{figure}

\subsection{Physical Parameters}
After data processing, we found the best fit orbital elements for the two newly discovered bodies using the fitting algorithm from \citet{2000AJ....120.3323B}. 
We computed refined astrometric positions for each wide field observation using the \textsc{WCSfit} software \citep{2017PASP..129g4503B}, which provides astrometric solutions referenced to the GAIA DR1 catalog \citep{GAIA_DR1}. This process included corrections for the effects of distortions on the DECam CCDs, as well as for chromatic terms from lateral color and differential atmospheric refraction.
The fit for \ThingOne\ used a series of 99 total exposures taken between December 21, 2013 and January 11, 2018. 
\ThingTwo\ was fit using a series of 39 observations taken between October 12, 2013 and December 26, 2017. 
For consistency, we also refit the orbital elements for 2013 UH${15}$ using the 10 available observations. 
The best-fit values, along with magnitudes and colors derived from DES photometric measurements, are presented in Table \ref{table:1}.
The colors for \ThingOne\ and \ThingTwo\ were computed by subtracting simultaneous (defined as taken on the same night, to account for systematic variability between nights that would lead to subtracted colors drifting when measurements from multiple nights were combined) measurements for each color and averaging across all measurements for an object. We ignore object rotation because each supernova field exposure sequence analyzed here consists of a set of $griz$ exposures and lasts just 14 minutes in total. This is much less than the typical rotation period of a TNO, so the color measurements at each epoch can be regarded as essentially simultaneous. The colors of \ThingOne\ and \ThingTwo\ are consistent with each other to within 1$\sigma$ for all four wavebands.

The orbital elements of \ThingOne\ and \ThingTwo\ are remarkably similar to each other and also to known object 2013 UH$_{15}$ \citep{2016MPEC....Q...40S}. These three objects have quantities $(a,e)$ varying by less than about 2 percent among the three bodies, and the remaining angles $(i,\omega,\Omega)$ are also similar: in particular, the orbits of \ThingOne\ and 2013 UH$_{15}$ appear to be nearly perfectly aligned. Figure \ref{fig:orbits} shows the orbits of the three bodies. The top panel shows all of the known ETNOs with semi-major axes $a>150$ AU and perihelion distances $q>35$ AU. A close-up of the orbits of the two new objects and their previously discovered counterpart is shown in the bottom panel. 

The orbits of these objects are not identical, and similar orbits do not prove a common origin. However, considered in the context that these objects reside in the outer Solar System, in a sparsely-populated region of
orbital parameter space ($a>150$ AU), such similar orbits are unusual and merit further study. 
The size ratios between the objects is also intriguing: assuming albedos of 10\%, \ThingOne, \ThingTwo, 
and 2013 UH$_{15}$ have sizes of 280 km, 110 km, and 120~km respectively. If all three objects are assumed to have the same albedo and density, the two smaller objects (2013 UH$_{15}$ and \ThingTwo) have about 15\% the mass of the largest object, \ThingOne.
\subsection{Observational Bias}
\begin{figure}[t!]
\epsscale{1}
  \begin{center}
      \leavevmode
\includegraphics[width=3.4in]{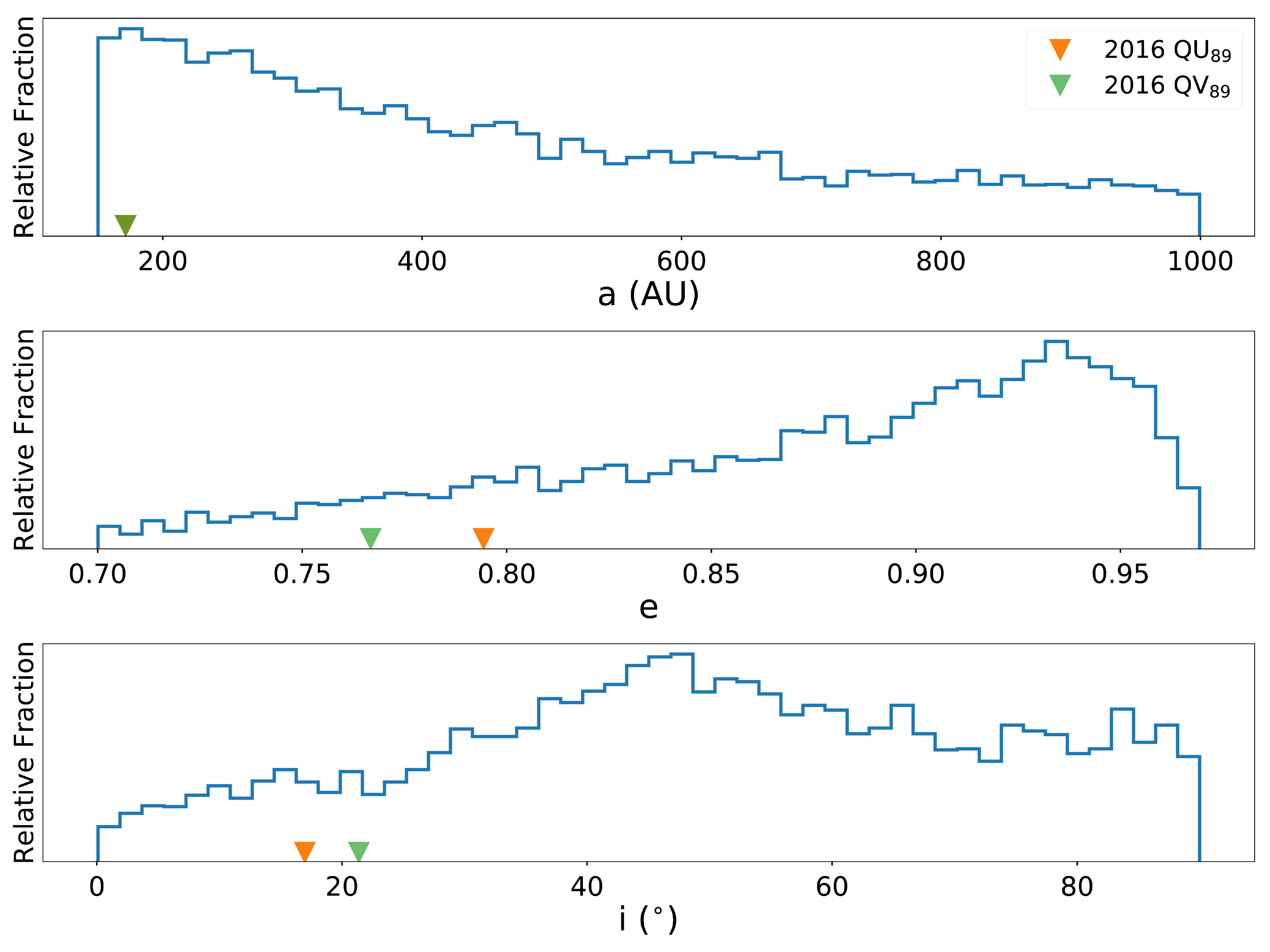}
\caption{DES selection function for objects with perihelion distances greater than 30 AU and semi-major axis greater than 150 AU and smaller than 1000 AU. The orbital elements of \ThingOne\ and \ThingTwo\ are denoted by triangles. Note that the two triangles are overlapping in the top panel due to the objects' similarity in semi-major axis. }
\label{fig:bias}
\end{center}
\end{figure}
Due to the small survey area (the relatively small sky-area three-square-degree supernova fields) where these objects were initially detected, it is possible that the apparent similarity in their orbital elements comes from bias in their detection locations. 
If this were the case, we would expect TNOs with the ($a,e,i$) of these objects to be more easily recovered by our detection pipeline. 
This would mean that the association between the objects is only due to a bias in the sensitivity of our pipeline. 

To test this, we simulate roughly 440,000 clones of objects dynamically similar to those considered in this work. The orbital elements of these clones are drawn from uniform distributions with semi-major axis $150 < a < 1000$ AU, perihelion distance $q>30$ AU, and inclination $i$ uniformly distributed between 0 and 180 degrees. Choosing this population of clones allows us to identify the selection biases of DES; this initial distribution is not meant to represent a realistic Kuiper Belt population. 
Using the computed orbits of these objects, we then determine which orbits could be detected in DES observations in the first three observing campaigns. First, we eliminate objects whose sky positions or apparent magnitudes made them unobservable by DES. Then, of the remaining clones, we test which would be observable and linkable by our pipeline. Clones are considered linkable in the supernova data if they are detectable in three or more DES exposures (on different nights), with neighboring exposures separated by 20 nights or fewer. 

A total of 6446 unique clones were determined to be observable using these criteria, and those surviving clones are plotted in Figure \ref{fig:bias}. 
These histograms show the distribution of instantaneous orbital elements for linked clones, as a fraction of the input uniform distributions. 
Since the resulting distributions are fairly flat and do not show any strong preference for orbital elements around the ($a,e,i$) values of \ThingOne\ and \ThingTwo, we can be confident that our detection of two objects with such similar orbits is not due to observational bias. 





\bigskip

\section{Dynamics and Orbital Similarities}
\label{sec:dynamics} 

\begin{figure*}[t!]
\epsscale{1}
  \begin{center}
      \leavevmode
\includegraphics[width=180mm]{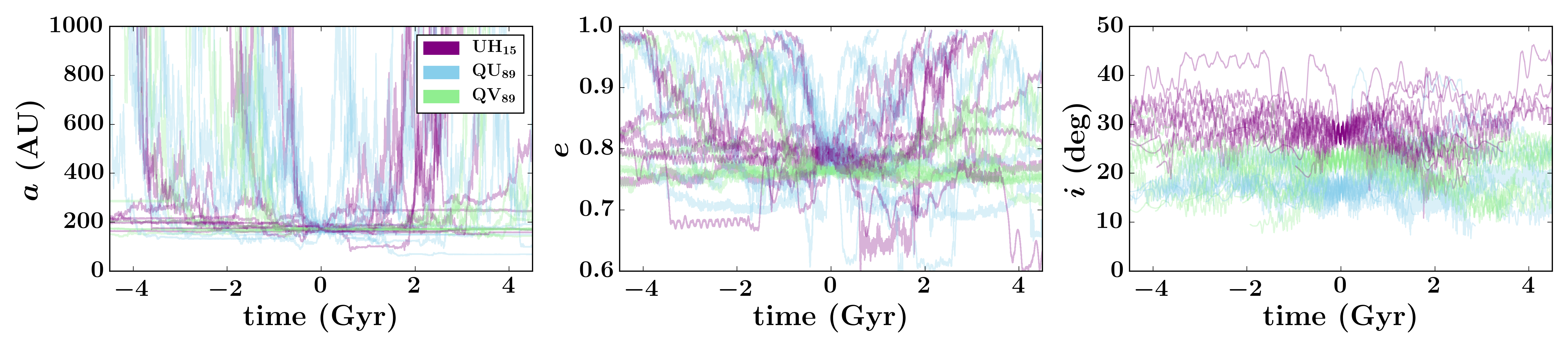}
\caption{Time-evolution of the semi-major axis, eccentricity, and inclination for representative clones of each object. The clones are initialized with sets of orbital elements that are statistically equivalent to the best-fit parameters. The clones of \ThingOne\ are shown in green, \ThingTwo\ in blue, and 2013 UH$_{15}$ in purple.} 
\label{fig:aeievolution}
\end{center}
\end{figure*}

In order to study the dynamics of these three objects, we employ a numerical model, as these ETNOs fall in the Neptune scattering region (with perihelion distances in the $q \sim 35-45$ AU range). As a result, their behavior can be driven by resonant effects \citep{2017CeMDA.127..477S, 2018AJ....155..260V} and their energy diffusion can be driven by close encounters with Neptune \citep{1987AJ.....94.1330D}. Because the evolution of these objects depends on the aforementioned  short-period effects, numerical simulations are an effective tool for understanding their dynamics (for comparisons of numerical and analytic treatments in related contexts, see, e.g., \citealt{2017MNRAS.468..549B, 2018MNRAS.474...20H}).
In this work, we use N-body code \texttt{Mercury6} to evaluate the evolution of these three objects in the presence of the four giant planets. We exclude the terrestrial planets and use a time-step of 20 days, with a hybrid symplectic and Bulirsch-Stoer (B-S) integrator. In each integration, we include test particle clones of each of the three TNOs, with orbital elements for each drawn from the covariance matrices resulting from our fits and corresponding to the values in Table \ref{table:1}. We integrate the orbital elements for each TNO to a common epoch before beginning the simulations. Simulations are run in two batches, one forwards and one backwards in time for 4.5 Gyr each. 
These simulations allow us to study the orbital evolution of these three objects (which we call the `triplet' objects) and assess their similarity. Before considering the potential association between the three TNOs, however, we explore their dynamics individually.

\subsection{Individual Dynamics}
\label{sec:individual}



\begin{figure}[t!]
\epsscale{1}
  \begin{center}
      \leavevmode
\includegraphics[width=90mm]{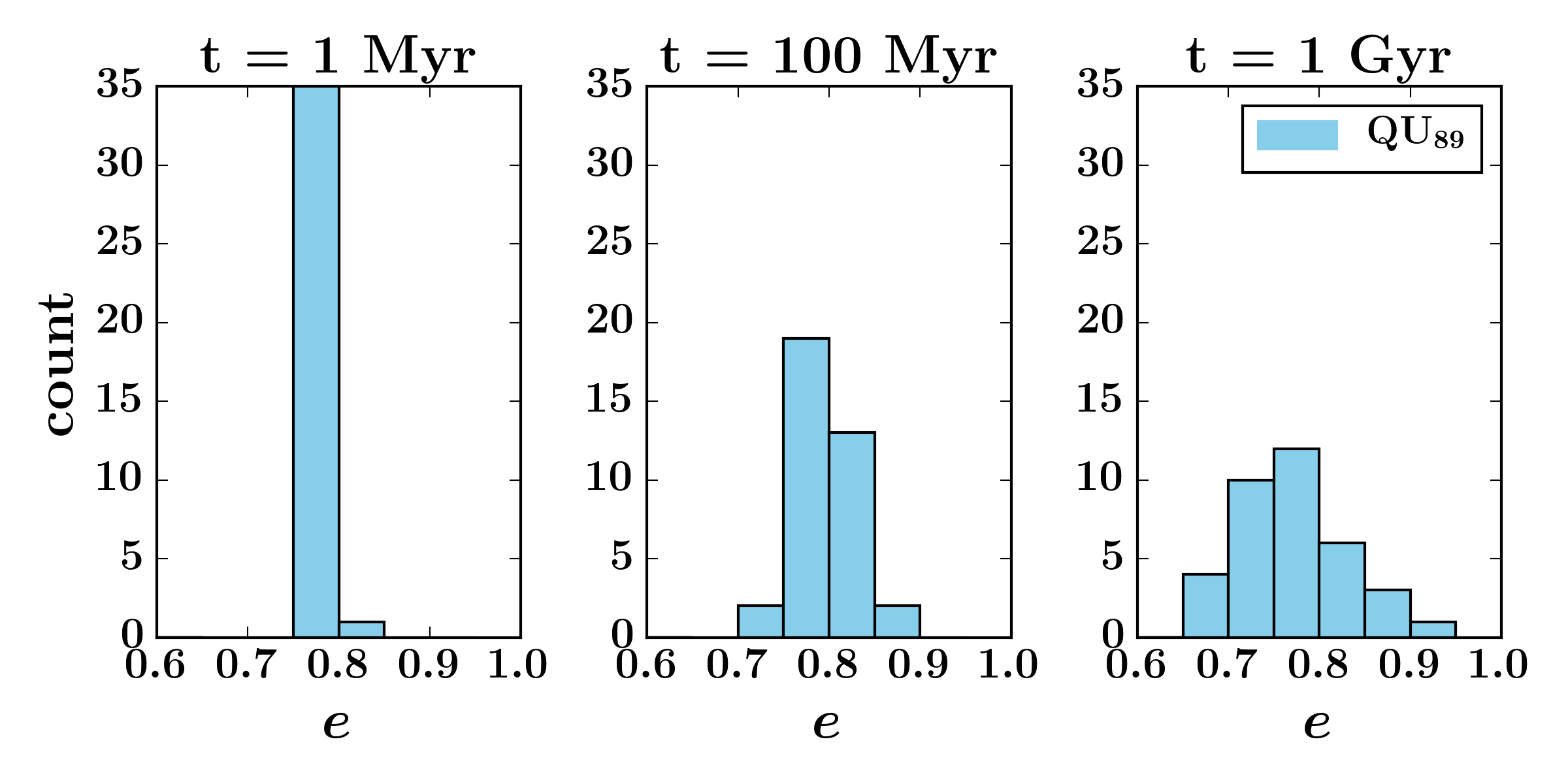}
\caption{Snapshots of the eccentricity distributions for the surviving clones of \ThingTwo \ at three epochs. At early times (1 Myr), the eccentricities of the clones are tightly centered around the best-fit value (left panel). At later times, this distribution spreads apart, with distributions shown for 100 Myr (middle panel) and 1 Gyr (right panel). The full time evolution of the distribution widths is shown in Figure~\ref{fig:sigmavstime}.} 
\label{fig:e_dist}
\end{center}
\end{figure}

Although all three of the triplet objects currently have long-period orbits, these ETNOs are not decoupled from the rest of the Solar System as their perihelia are bound to Neptune. This effect is evident in Figure \ref{fig:aeievolution}, which shows the evolution of the semi-major axis, eccentricity, and inclination of a representative subset of triplet object clones. Due to scattering interactions with Neptune, the semi-major axis of the triplets changes rapidly and their eccentricities grow. As a result, not all clones survive the entire Solar System lifetime. In fact, only $36 \%$ of the clones of \ThingTwo \ survive the full $4.5$ Gyr simulation; similarly, only $45 \%$ of the clones of 2013 UH$_{15}$ survive. Interestingly, the clones of \ThingOne\ are significantly more stable, with $89 \%$ of clones surviving, possibly because of
its larger perihelion distance. The stability fractions for the backwards integrations are similar, as expected, with $29 \%$ clones surviving for \ThingTwo, $47 \%$ for 2013 UH$_{15}$, and $82 \%$ for \ThingOne.

To characterize this behavior in more detail, we consider how the orbital elements of clones of a single object change over time. For example, in Figure \ref{fig:e_dist}, we present several snapshots of the eccentricity distribution of the surviving clones of \ThingTwo. Initially, this distribution is tightly centered around the best-fit value. As the system evolves, the eccentricity distribution spreads apart, as shown by the three panels in Figure \ref{fig:e_dist}. In order to further characterize this behavior, we consider the time-evolution of the distribution width of several parameters. 

\begin{figure}[t!]
\epsscale{1}
  \begin{center}
      \leavevmode
\includegraphics[width=85mm]{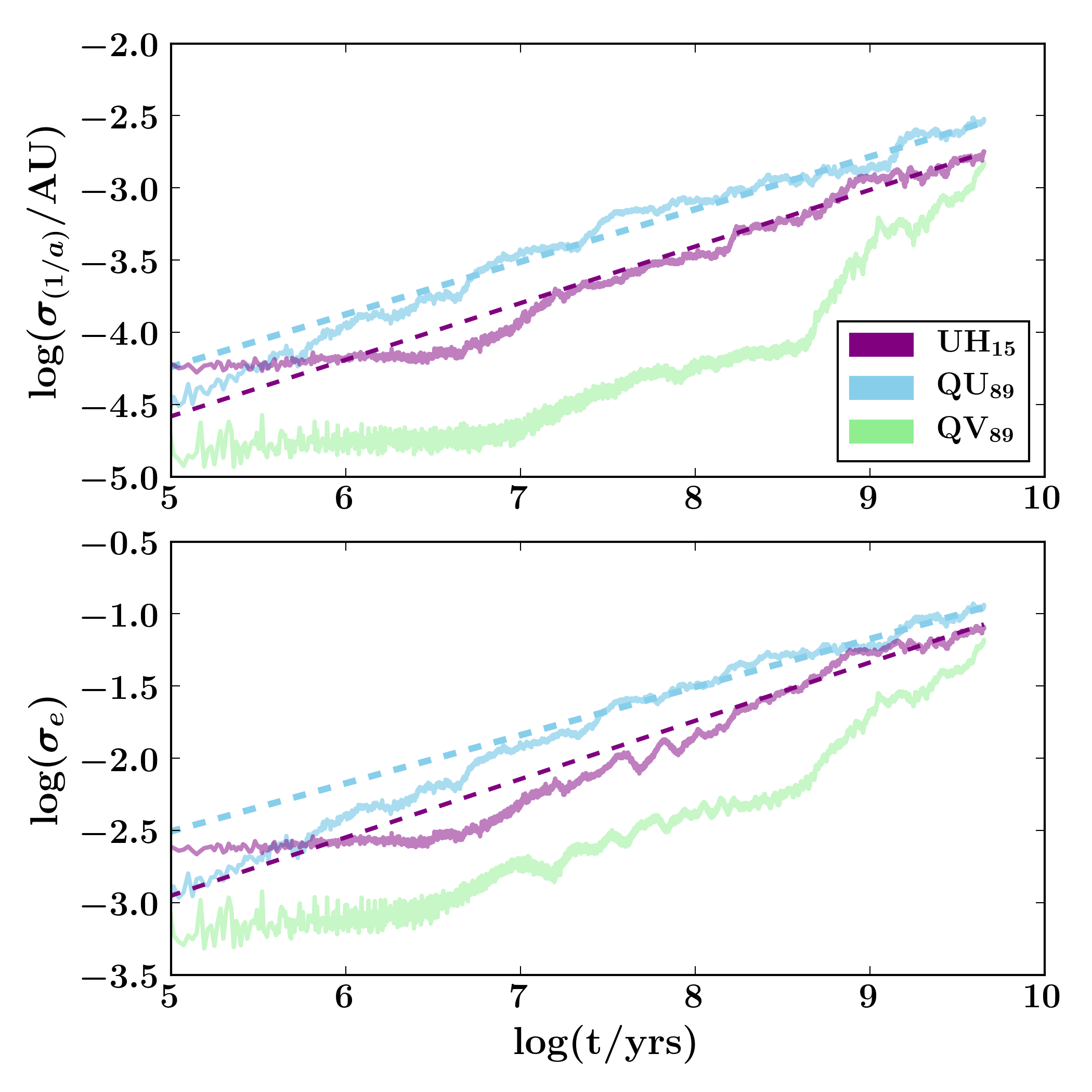}
\caption{The time evolution of the widths $\sigma$ of the distributions of the orbital parameters $1/a$ and $e$. Here $\sigma$ is defined to be the standard deviation of the distribution of a given orbital element for a population of clones, which have been integrated numerically forward in time. The dashed lines show linear fits to the curves (in the log-log plot) for the time span 0.1 Myr -- 4.5 Gyr for all three objects. The slopes for \ThingTwo \ and 2013 UH$_{15}$ are $m \approx 0.4$, which is somewhat shallower than the benchmark value $m=0.5$ expected for diffusive behavior. \ThingOne \ does not appear to fit the linear trend well.} 
\label{fig:sigmavstime}
\end{center}
\end{figure}

\begin{figure}[t!]
\epsscale{1}
  \begin{center}
      \leavevmode
\includegraphics[width=85mm]{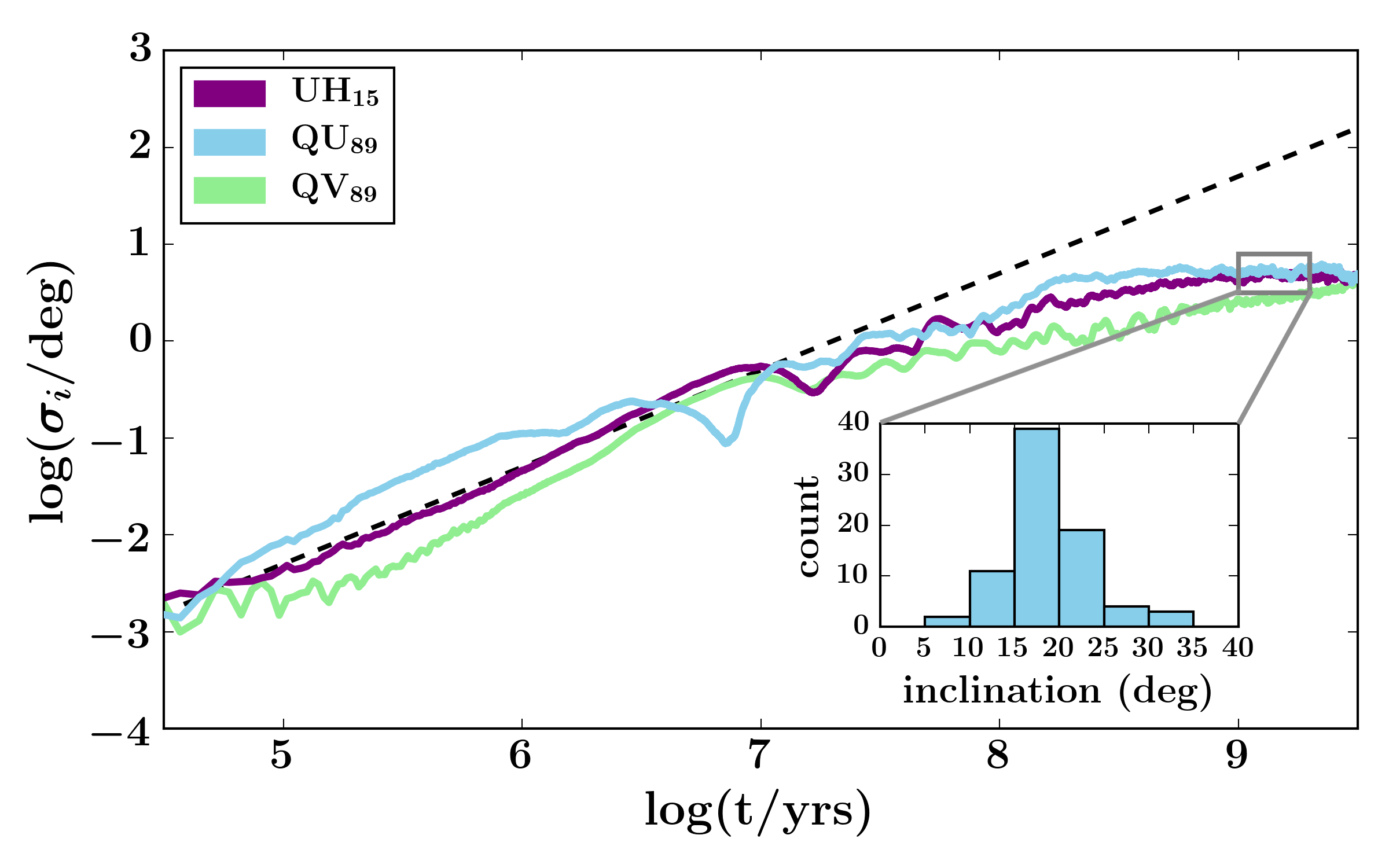}
\caption{The time evolution of the width $\sigma$ of the inclination distribution, where $\sigma$ is defined as in Figure \ref{fig:sigmavstime}. The guiding dashed line has a slope of 1 on this log-log plot, showing that initially the width of the distribution grows, but soon plateaus to a roughly constant value. This indicates that the inclination distribution has reached a point after which it does not significantly change; this distribution is shown in the inset plot for \ThingTwo.} 
\label{fig:inc_sigma}
\end{center}
\end{figure}

In particular, we focus on the distribution width of the parameter $1/a$ (which scales with orbital energy), orbital eccentricity $e$, and inclination $i$. We compute the standard deviation $\sigma$ of the parameter of interest among all surviving clones of each object at every time step. The resulting time evolution of $\sigma$ is shown in Figures \ref{fig:sigmavstime} (for $1/a$ and $e$) and \ref{fig:inc_sigma} (for $i$) on a log-log plot.

The initial distribution of clones of each orbital element resembles a Gaussian, centered at the best fit value of the parameter. In Figure \ref{fig:sigmavstime}, we show the behavior of $\sigma$ for the parameters $1/a$ and $e$. For the first few million years, the distribution hardly changes, as the orbits of the clones do not evolve on such short time scales. Over the span of four billion years, however, the spread in the clones increases. Linear fits to the blue (\ThingTwo) and purple (2013 UH$_{15}$) $\sigma$-curves find a slope of $m \approx 0.4$, which indicates that these clones spread apart in a sub-diffusive manner. Note that diffusion can be described as a random walk, so that quantities should increase with time $\propto t^{1/2}$. One might expect the distribution of orbital elements to behave similarly, as a random walk in phase space, due to scattering interactions with Neptune (subject to an absorbing boundary at $e=1$, where objects are ejected from the Solar System). If this were, in fact, a diffusion process, we would expect to see a straight line with slope $m=0.5$ on the log-log plot. However, it is evident that the observed behavior is not a pure random walk, as the time evolution of the $\sigma$ values are somewhat slower. 

The behavior of the green $\sigma$-curve (\ThingOne) differs from the two others. It appears that the dynamics of \ThingOne \ are not as dominated by Neptune scattering as are the other two objects. This trend is evident in the much higher number of surviving clones of \ThingOne \ cited above. The reason for this difference is likely due to the higher perihelion distance of \ThingOne \ as compared to the other objects. With a perihelion of $q \approx 40$ AU, this object is more detached from Neptune and experiences fewer (and less disruptive) close encounters. In addition, the evolution of \ThingOne\ may be affected more strongly by mean-motion resonances with Neptune, a possibility that we discuss in Section \ref{sec:neptune_res}.

In Figure \ref{fig:inc_sigma}, we show the time-evolution of the width of the inclination distribution for the three objects. Initially, the width grows, roughly following a line with a slope of unity on the log-log plot, but appears to saturate on a time scale of order 0.1 Gyr, indicating that the inclination distribution remains approximately the same for the rest of the evolution. Notice that this last decade and a half in time corresponds to most of the age of the Solar System. This final inclination distribution is shown for \ThingTwo \ in the inset plot; most of the distribution falls in the range 15 -- 25 degrees. 

These results (in particular, Figures \ref{fig:aeievolution}, \ref{fig:sigmavstime}, and \ref{fig:inc_sigma}) show that the future orbits of these objects diverge steadily on relatively rapid timescales. As a result, any similarity in their current-day orbits must be due to either a recent event (such as fragmentation or a  binary dissociation event), the past attainment of an orbital resonance that would force the objects to maintain similar orbits, or by random chance. Taking this dynamical evolution into account, we now consider possible explanations for the similarity of the orbits of these three objects. In the next section, we evaluate the probability that the orbits of these objects are similar by random chance. 

\subsection{Comparison of Orbits}
\label{sec:comparison}



To evaluate the likelihood that the orbits of these objects are similar merely by coincidence, we compare the orbital similarity of the triplets to that of randomly chosen groups of three objects drawn from several control distributions.  

First, we define a dimensionless distance metric to characterize a set of three objects. This function $d(t)$ is taken to be the sum of the pairwise distances in the space of orbital elements. Specifically, we use the sum of Euclidean metrics acting on the scaled elements $a$, $e$, and $i$ for each pair of TNOs, i.e.,  
\begin{equation}
d(t) = \sum_{j\ne k} \Bigg[ \left(\frac{a_j(t) - a_k(t)}{\frac{1}{2}(a_{j_0} + a_{k_0})} \right)^2 + \qquad
\label{eq:distance}
\end{equation}
$$
\qquad 
\left(\frac{e_j(t) - e_k(t)}{\frac{1}{2}(e_{j_0} + e_{k_0})}\right)^2  + \left(\frac{i_j(t) - i_k(t)}{\frac{1}{2}(i_{j_0} + i_{k_0})}\right)^2 \Bigg]\,,
$$
where $j$ and $k$ denote the objects, and the subscript $0$ denotes the initial orbital elements of the objects.


\begin{figure*}[t!]
\epsscale{1}
  \begin{center}
      \leavevmode
\includegraphics[width=180mm]{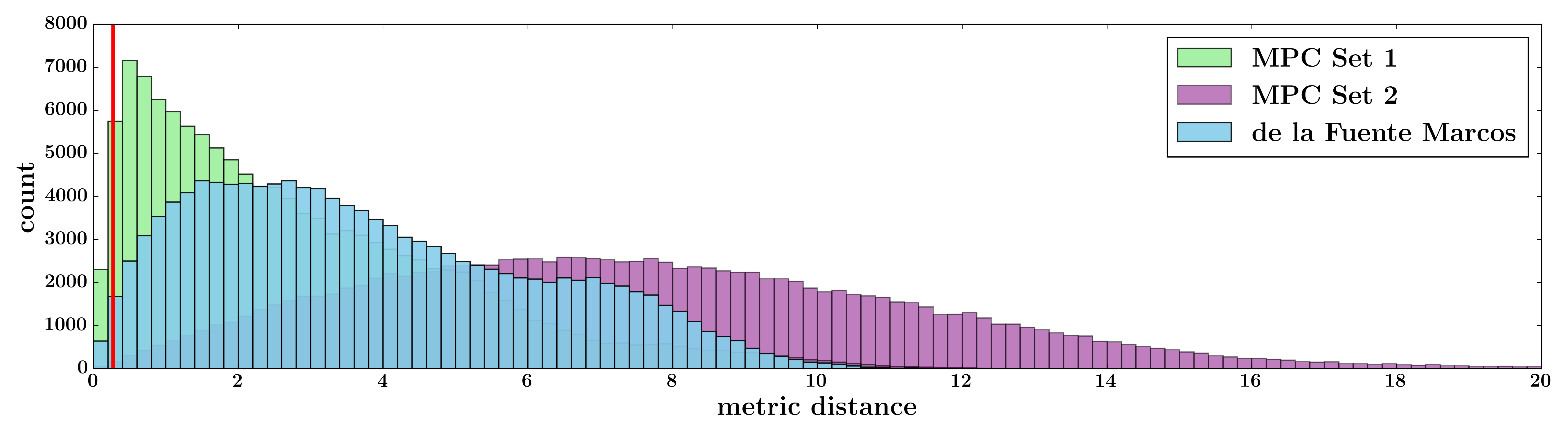}
\caption{A plot showing the current metric distance between the objects \ThingOne, \ThingTwo, and 2013 UH$_{15}$ (red vertical line) as compared to the distances between random sets of three objects drawn from several distributions. The green population (MPC Set 1) contains all of the TNOs reported to the Minor Planet Center with $a > 50$ AU, $e > 0.45, q > 30$ AU; the purple population (MPC Set 2) contains \textit{all} of the TNOS reported to the Minor Planet Center; and the blue population is the test particle distribution from \citet{2017Ap&SS.362..198D}, where $150 < a < 800$ AU, $0.7 < e < 0.95$, $i < 55^{\circ}$, and $30 < q < 90$ AU. As we can see, the triplet objects discussed in this work are currently closer together in phase space than most objects, for all three distributions considered. For all MPC objects, we used barycentric orbital elements with solution epoch 2458200.5 JD, which were converted from heliocentric orbital elements coming from an MPC file downloaded on 9-Jun-2018.}
\label{fig:MPC_real_comp}
\end{center}
\end{figure*}

Next, we choose a clone of each of  \ThingOne, \ThingTwo, and 2013 UH$_{15}$ from the measured orbit posteriors and compute the above metric distance at the current epoch.
Repeating this process for $125,000$ distinct combinations of clones, we compute the current average distance between our triplet.

From this analysis, we have a measure of the current-day orbital similarity of these three objects. 
To explore whether this distance is remarkable among the larger set of objects in the outer Solar System, we compare the current average distance of the triplet objects to several different representative distributions of TNOs, shown in Figure \ref{fig:MPC_real_comp}. Each of the distributions was created by drawing 125,000 sets of three objects from the specified population of TNOs and computing their current day metric distance. The green distribution (MPC Set 1) contains all of the TNOs reported to the Minor Planet Center with $a > 50$ AU, $e > 0.45, q > 30$ AU; the purple population (MPC Set 2) contains \textit{all} of the distant objects reported to the Minor Planet Center (we note that this includes Neptune trojans); and the blue population is the uniform test particle distribution from \citet{2017Ap&SS.362..198D}, with $150 < a < 800$ AU, $0.7 < e < 0.95$, $i < 55^{\circ}$, and $30 < q < 90$ AU. 

The red vertical line in Figure \ref{fig:MPC_real_comp} shows the current day average distance between our triplet objects. It is clear that this red line falls on the far left of all three TNO distributions considered. For MPC Set 1, 4023 members of the 125,000 sets in the control population, or 3.2\%, were more similar than the observed triplet. For MPC Set 2, 96 of 125,000 sets, or 0.077\%, were more correlated; finally, for the distribution drawn from \citet{2017Ap&SS.362..198D}, 1137 out of 125,000 sets, or 0.9096\%, were more correlated than the triplet. The robust nature of this result leads us to conclude that the current day similarity of \ThingOne, \ThingTwo, and 2013 UH$_{15}$ is highly unusual, even when compared to a variety of background populations. 

Considering the distance between objects, as outlined above, may provide an effective method of identifying other triplet sets from the whole population of KBOs. Toward this end, we apply our analysis to the dwarf planet Haumea and its associated family members. Although a full characterization of the dynamics of the Haumea system is beyond the scope of this work, it provides a useful test of our method. The Haumea system contains two moons, which are known from previous literature to be associated \citep{2007Natur.446..294B, 2008AJ....136.1079L, 2008ApJ...684L.107S, 2009AJ....137.4766R, 2009ApJ...700.1242S, 2010ApJ...714.1789L, 2012Icar..221..106V}.

For this test, we compute the current day metric distance among all possible sets of three objects drawn from the Haumea family. We find that the Haumea triplets have distances directly comparable to the values computed for our triple system. 
The metric distance for triplets of objects drawn from the Haumea family ranges from 0.0024 to 1.28, showing that the correlated Haumea family is much more similar by our metric than the general control populations considered in Figure \ref{fig:MPC_real_comp}.
One should keep in mind, however, that the dynamical environment of the Haumea family is significantly different from that of the new triplet. Jupiter has a much larger effect on the evolution of Haumea due to its closer proximity. Moreover, it is possible that the break-up of the original dwarf planet into its present family members was caused by rotational fission \citep{2010A&A...511A..72S, 2012MNRAS.419.2315O}. As a result, comparisons between these two sets of objects can only be made at the order of magnitude level. 

The main conclusion from the distance analysis of this section is that the triplet objects (\ThingOne, \ThingTwo, and 2013 UH$_{15}$) are statistically more correlated than the general population of TNOs. Using the distance metric from equation (\ref{eq:distance}), we compare their orbital elements to those derived for three different control populations. For each population considered, the triplet objects are among the most correlated; however, we cannot exclude that the similarity in overall orbits is due to random chance. 


\begin{figure*}[t!]
\epsscale{1}
  \begin{center}
      \leavevmode
\includegraphics[width=6in]{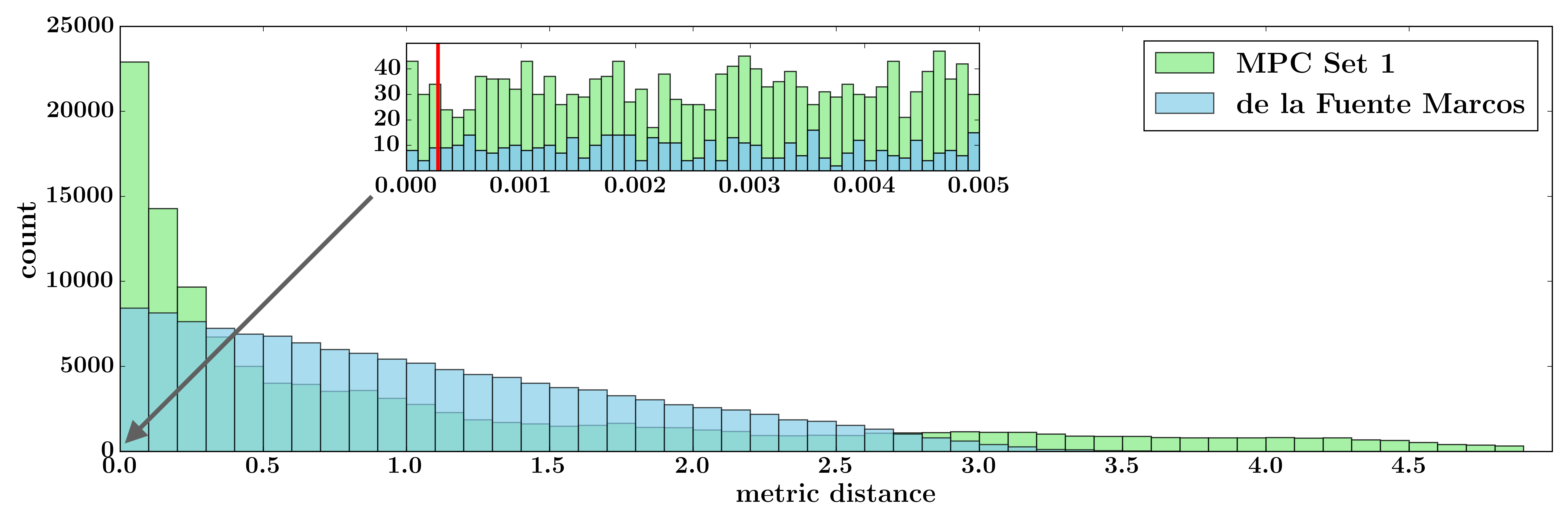}
\caption{A plot showing the current metric distance for semi-major axis only (Eq. \ref{eq:sma}) for the three objects considered in this work, as compared to the distances between random sets of three objects drawn from our control distributions (MPC Set 1 contains all of the TNOs reported to the Minor Planet Center with $a > 50$ AU, $e > 0.45, q > 30$ AU, the de la Fuente Marcos (dlFM) distribution is the test particle distribution from \citealt{2017Ap&SS.362..198D}, where $150 < a < 800$ AU, $0.7 < e < 0.95$, $i < 55^{\circ}$, and $30 < q < 90$ AU). The inset shows a zoom-in on the left side of the distribution, with the red vertical line indicating the current distance among our three triplet objects.}
\label{fig:MPC_aonly}
\end{center} 
\end{figure*}

Since the similarity between the semi-major axes of the triplets is the most unusual out of the other parameters, we consider one more test. Recomputing the metric we used previously for only semi-major axis, as follows,
\begin{equation}
d_{a}(t) = \sum_{j\ne k} \left(\frac{a_j(t) - a_k(t)}{\frac{1}{2}(a_{j_0} + a_{k_0})} \right)^2 
\label{eq:sma}
\end{equation}
yields the histograms in Figure \ref{fig:MPC_aonly}. In this figure, we compare MPC Set 1 and the de la Fuente Marcos distribution, which have similar eccentricities to our objects.

We find that compared to the present day distance between our triplet objects, 0.08\% of trials in Set 1 are more similar than our triplet and 0.01\% of trials in the dlFM set are more similar. We exclude Set 2 because it includes Neptune trojans, which we expect to have fixed, identical values of semi-major axis. 

The analysis based on semi-major axis alone describes these objects as more remarkable than the full metric using all orbital elements. 
The existing population of TNOs with semi-major axis values greater than 150 AU remains small, with these three objects representing roughly 10\% of the known total at the time this paper was written. As such, it seems striking that these three would reside in the same $\sim$2 AU of such a large parameter space.
One explanation that can force objects to attain particular values of semi-major axis (but affect the objects' eccentricities and inclinations less directly) is orbital resonance, which we discuss in the next section.



\vskip0.8truecm

\section{\ThingOne, \ThingTwo, and 2013 UH$_{15}$ \\
as Resonant Objects}
\label{sec:resonance}

To explore the possibility that the semi-major axis similarity of the triplets is explained by orbital resonances, we evaluate the likelihood of these objects falling into resonances with other Solar System bodies, specifically, Neptune and the proposed new Solar System member Planet Nine. We find that the resonant dynamics in each of these two cases are distinct. 

\subsection{Resonances with Neptune}
\label{sec:neptune_res}

Thus far, we have discussed only the scattering interactions that these three objects experience due to Neptune. In theory, however, these ETNOs, although distant, may be affected by mean-motion resonances with Neptune. In fact, recent work by \citet{2018AJ....155..260V} has identified two objects with semi-major axes of 129.8 AU and 129.9 AU as living in the 9:1 resonance with Neptune. These two objects are currently the most distant TNOs known to reside in Neptune resonances. 

Inspired by this finding, we perform a similar analysis on the clones of our three objects. We identify intervals of time in which the period ratio between a clone and Neptune is approximately constant, and then test resonances up to 29th order, considering resonances that fall into the period ratio interval $(P_{\rm Nep}/P_{TNO}) \pm 0.1$. For each resonance argument, we generate plots of the time-evolution and identify intervals of libration by finding low-point-density regions in the plots. 

\begin{figure}[t!]
\epsscale{1}
  \begin{center}
      \leavevmode
\includegraphics[width=85mm]{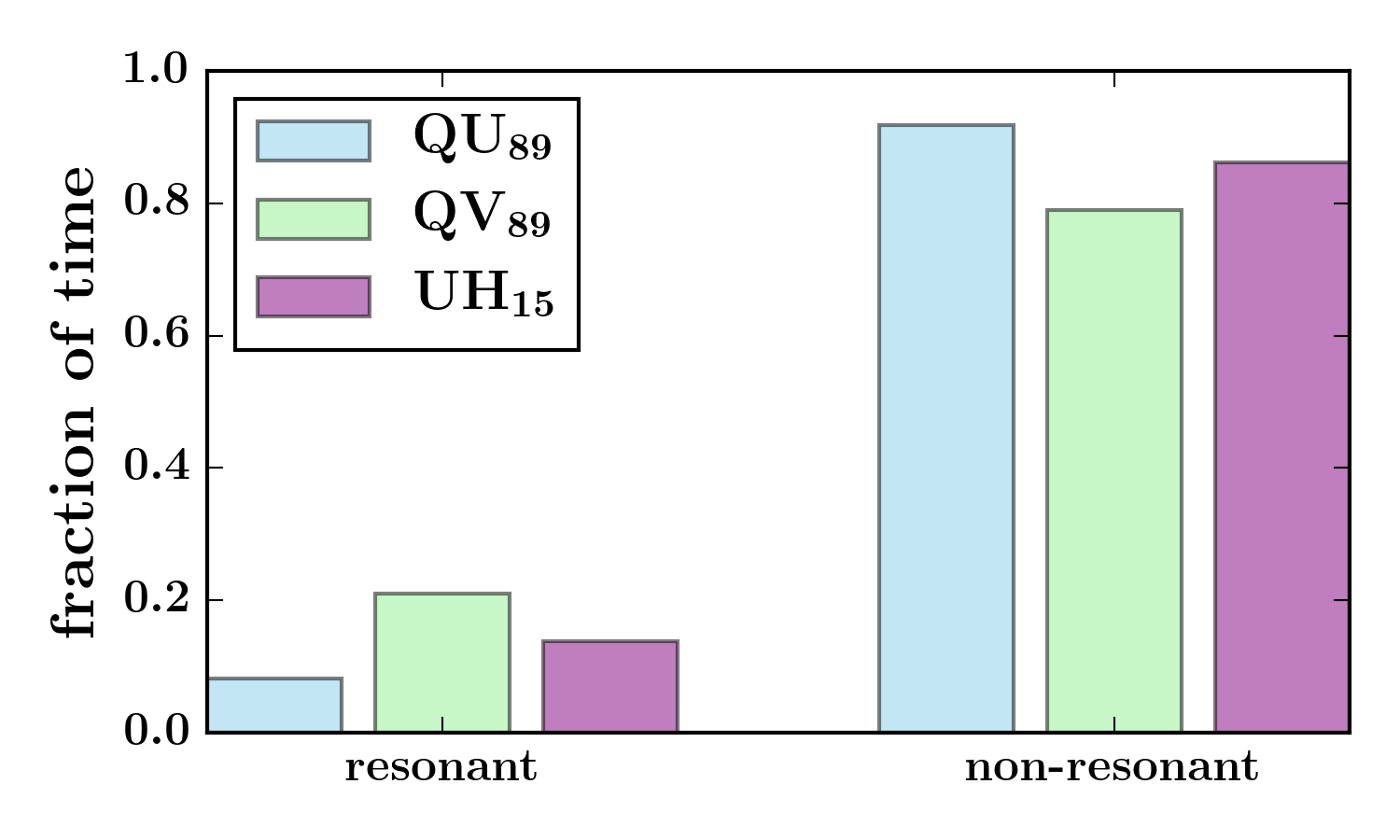}
\caption{The fraction of time each object considered in this work spends in a Neptune resonance in a 4.5 Gyr backwards integration. To compute the fraction of time, we sum the total time spent in resonance by all of the clones of an object and divide by the total survival time of those clones. \ThingOne \ is the most resonant object of the three, spending about $20\%$ of its time in resonance.} 
\label{fig:res_ratios}
\end{center}
\end{figure}

The output of the above analysis allows us to identify the time intervals during which a clone is securely librating in a resonance. Interestingly, we find that these three objects often experience resonant interactions with Neptune, which usually last tens of millions of years. Summing over the evolution of all of the clones, we find that \ThingOne \ is resonant most often, with $21 \%$ of the total integration time spent in resonance, while \ThingTwo \ and 2013 UH$_{15}$ spend $8 \%$ and $14 \%$ of the time in resonance, respectively (see Figure \ref{fig:res_ratios}).

\begin{figure}[t!]
\epsscale{1}
  \begin{center}
      \leavevmode
\includegraphics[width=85mm]{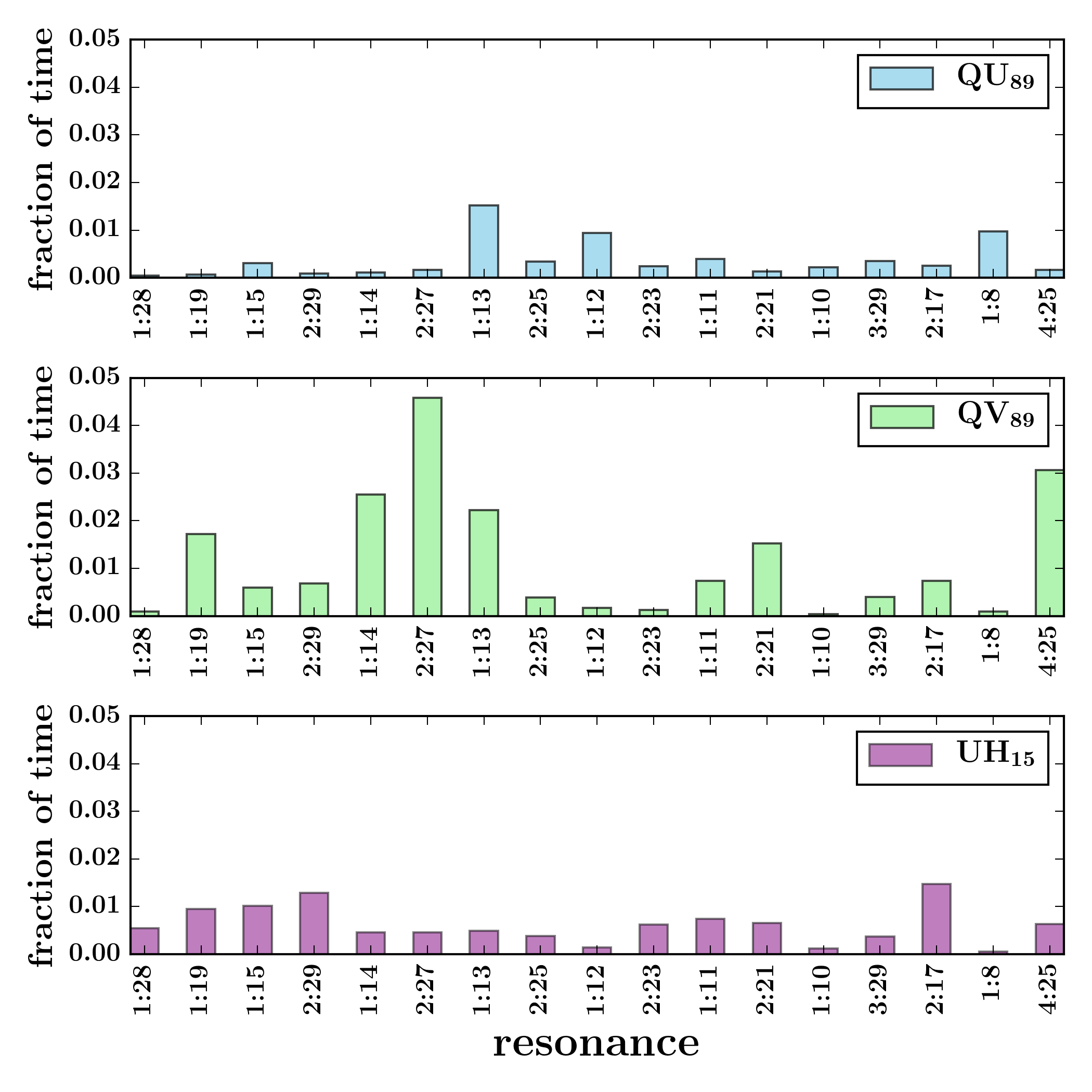}
\caption{Histograms showing the time spent in resonant configurations with Neptune. The bins on the $x$-axis represent the union of the ten most populated resonances for each object. The fraction of time spent in a resonance is computed by summing over all clones of an object, using the same procedure as in Figure \ref{fig:res_ratios} (see text).} 
\label{fig:res_hist}
\end{center}
\end{figure}

\begin{figure}[t!]
\epsscale{1}
  \begin{center}
      \leavevmode
\includegraphics[width=85mm]{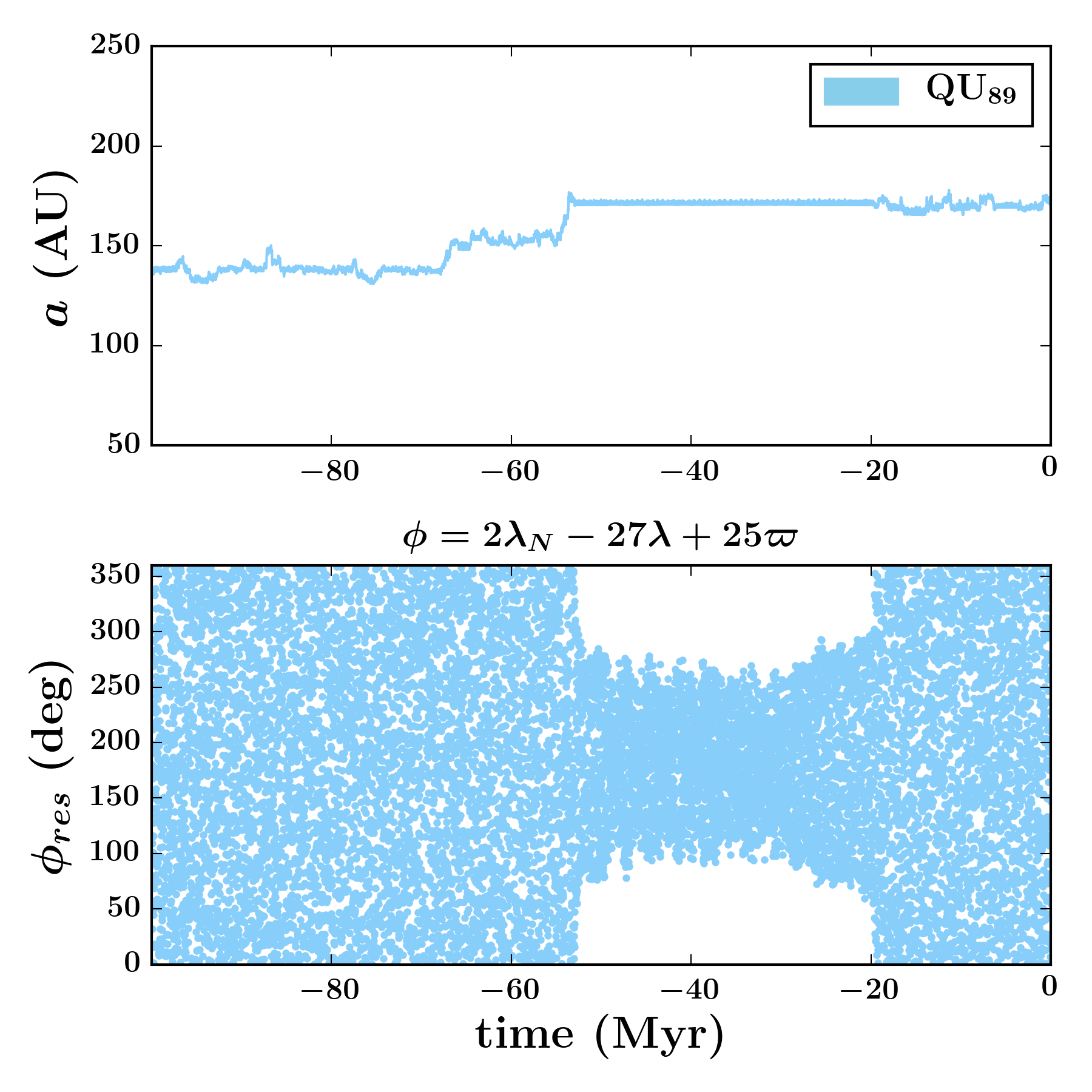}
\caption{An example of the $2$:$27$ resonance in the backwards integrations for \ThingTwo. The top panel shows the semi-major axis evolution and the bottom panel plots the appropriate resonance argument over the same time interval. The time spent in resonance is brief, less than 20 Myr. Note that time interval for the libration of $\phi_{res}$ (bottom) clearly corresponds to the regime of constant semi-major axis evolution (top).} 
\label{fig:res_example}
\end{center}
\end{figure}

The most populated resonances for these three objects are shown in Figure \ref{fig:res_hist}. Here we take the ten most populated resonances for each of the three objects and then consider their union. Since the most populated resonances are not the same for all three bodies, we plot seventeen resonances in total. As expected, most of these resonances are quite high order. For instance, the most populated resonance for \ThingOne \ is the $2$:$27$ commensurability, which corresponds to a semi-major axis of about $a=170$ AU, close to its current day value. An example of such a resonant instance is shown in Figure \ref{fig:res_example}.

It is important to note, however, that these intervals of resonance are transient. None of the clones of these objects are currently in resonance, so they cannot be classified as resonant TNOs according to the standards of \citet{2008ssbn.book...43G}. Nonetheless, the fact that these objects are likely to have been in Neptune resonances in the past provides an additional explanation for their orbital similarity. As shown in Figure \ref{fig:res_example}, even after true resonance is lost, a TNO may reside in a semi-major axis value close to the (previously) resonant value for some time. It is possible that these objects were recently in the same Neptune resonance and are presently evolving away from that state, explaining their current similarity in semi-major axis.

\subsection{Resonances with Planet Nine}
\label{sec:planetnine} 



Recently, observed correlations in argument of perihelion \citep{st14} and then in longitude of perihelion \citep{BB16} among the most distant TNOs have led to the proposal of a ninth Solar System planet, often called Planet Nine or Planet X. 
If this proposed planet exists, it would also provide a network of mean-motion resonances in which TNOs could reside. In this section, we consider how the three ETNOs discussed in this work fit into the context of the Planet Nine hypothesis.

As shown in Figure \ref{fig:orbits}, the orbital orientation (as defined by the longitude of perihelion, $\varpi$) of these three objects appears to be anti-aligned with the proposed orientation of the Planet Nine orbit. This clustering is thus consistent with the orbits of the extreme TNOs first used to infer the existence of Planet Nine and predict its orbit. However, these objects have semi-major axes ($a\sim 171- 173$ AU) significantly shorter than those of TNOs thought to be dominated by Planet Nine-induced dynamics. \citep[Note that different models for Planet Nine imply different inner boundaries (in $a$) for the region affected by the putative planet. The most recent analysis,][shows that objects with $a>250$ AU are likely to reside in the regime influenced by Planet Nine]{BatMorb}. As a result, despite the consistent orbital properties of these three ETNOs, it is unclear whether these objects could be part of the securely anti-aligned population. 

In addition to maintaining the apsidally anti-aligned population, Planet Nine has been shown to generate TNO orbits that are apsidally aligned with the orbit of Planet Nine, and others that experience apsidal circulation \citep{BB16, BatMorb, summerwork}. In the presence of Planet Nine, these three objects of interest could fall into either the anti-aligned or circulating categories. In this section, we discuss the dynamics of these TNOs in the presence of representative Planet Nine candidates and consider possible mean-motion resonances.



To differentiate between the behaviors outlined above, we define the difference between the longitude of perihelion of Planet Nine and a TNO as
\begin{equation}
\Delta \varpi \equiv \varpi - \varpi_9,
\end{equation} 
where $\varpi$ is the longitude of perihelion of the TNO and the subscript $9$ denotes the orbital elements of Planet Nine. Previous studies have shown that objects that are heavily influenced by Planet Nine do not retain a constant $\Delta \varpi$ \citep{2016A&A...590L...2B, BatMorb}. Instead, TNOs aligned with Planet Nine experience libration around $\Delta \varpi \sim 0^{\circ}$, anti-aligned objects librate about $\Delta \varpi \sim 180^{\circ}$, and circulating objects have $\Delta \varpi$ that circulates (by definition) through all values in $[0^{\circ}, 360^{\circ}]$.

To determine which class of behavior a given TNO experiences, it is not enough to know only its current $\Delta \varpi$; we need to integrate its orbit forward in the presence of Planet Nine and the current Solar System, and analyze the time-evolution of its $\Delta \varpi$. 
To evaluate the effect of Planet Nine on the orbital similarity of these three ETNOs, we use numerical simulations similar to those run in the previous section, but now including Planet Nine. 


Different incarnations of Planet Nine lead to distinct behaviors for these three ETNOs. Since the orbital elements of Planet Nine are not yet well constrained, we vary its orbit over a range of parameter space (as in \citealt{Becker}) to examine each class of possible interactions between Planet Nine and the three ETNOs under consideration. As shown below, we examine each of the possible dynamical classes that these objects could belong to by adjusting Planet Nine's orbit, and focus on the question of the orbital similarity of the triplet objects in each of these cases. 

It is important to note that an exhaustive survey of parameter space, as well as determining the exact limits of the Planet Nine parameters that cause each type of behavior considered are interesting questions, but beyond the scope of this work. Instead, we consider two test cases to illustrate --- but not comprehensively study --- the regimes of possible behavior.  

In Section \ref{sec:antialigned}, we study a Planet Nine candidate with semi-major axis $a_9$ = 315 AU, an eccentricity $e_9$ = 0.5, and an inclination $i_9$ = 20 degrees. This orbit lies near the inner edge of the parameter space proposed for Planet Nine and is found to induce apsidal anti-alignment in the triplet orbits. In Section \ref{sec:circres}, we then consider a Planet Nine candidate with orbital elements $a_9$ = 505 AU,  eccentricity $e_9$ = 0.5, and inclination $i_9$ = 20 degrees. This orbit lies near the center of the proposed parameter space and allows for the triplets to be apsidally circulating.

For the numerical work carried out in this section, all simulation parameters (time-step, etc) are identical to those of earlier integrations, with the exception of the introduction of a massive body in the form of the one proposed within the framework of the so-called Planet Nine hypothesis.

\subsubsection{ETNOs with Anti-Aligned $\Delta \varpi$}
\label{sec:antialigned}

Independent of considerations of their orbital similarity, these three objects are currently roughly anti-aligned with the orbit of the proposed Planet Nine, with a longitude of perihelion $\varpi = \omega + \Omega$ that orients their orbits between Sedna and 2012 VP$_{113}$ (Figure \ref{fig:orbits}). 
As mentioned above, however, an instantaneous anti-aligned $\Delta \varpi$ is not sufficient to sort these objects into the anti-aligned class, as they could just be opportunely observed ETNOs whose $\Delta \varpi$ are truly circulating. In the presence of an appropriately chosen Planet Nine, however, these objects do experience librations in the offset of longitude of perihelion around $\Delta \varpi \sim 180^{\circ}$. 

Given the parameters of Planet Nine, it is possible to approximate the semi-major axis threshold at which the anti-alignment of the distant Kuiper Belt begins \citep{Becker, BatMorb}. That is, depending on the orbit of Planet Nine, we can estimate which outer Solar System objects are expected to experience librations about $\Delta \varpi \sim 180^{\circ}$. Assuming a fixed mass for Planet Nine, the location of this threshold is determined (in part) by the perihelion distance of Planet Nine, and is thus a function of $a_9$ and $e_9$. Using these estimates as a guideline, we run a 1 Gyr integration of the clones of the three ETNOs in the presence of the four gas giants and a Planet Nine with orbital elements $a_9 = 315$ AU, $e_9 = 0.5$, $i_9 = 20^{\circ}$, $\omega_9 = 150^{\circ}$, $\Omega_9 = 120^{\circ}$, and mean anomaly $M_9 = 180^{\circ}$, which are parameters that fall in the anti-aligned region for these objects. Indeed, these simulations show that the ETNOs experience libration in the offset of longitude of perihelion around $\Delta \varpi \sim 180^{\circ}$. 

Due to their anti-alignment, these objects reside in an orbit-crossing region with Planet Nine. In order to avoid close encounters and consequent instability, such ETNOs are often in mean-motion resonances with Planet Nine \citep{2016ApJ...824L..22M,sarah}. In fact, they exhibit a more complicated behavior now known as ``resonance hopping'' \citep{Becker, BatMorb, 2017arXiv171206547H}, where the objects are often locked into mean-motion resonance, but transition into different resonances over the age of the system. It is important to note that mean-motion resonances, which correspond to libration of resonant arguments $\phi$ (e.g., $\phi=(p+q)\lambda_9-p\lambda-q\varpi_9$), are distinct from apsidal resonances, which correspond to libration of $\Delta{\varpi}$. 

\begin{figure}[t!]
\epsscale{1}
  \begin{center}
      \leavevmode
\includegraphics[width=85mm]{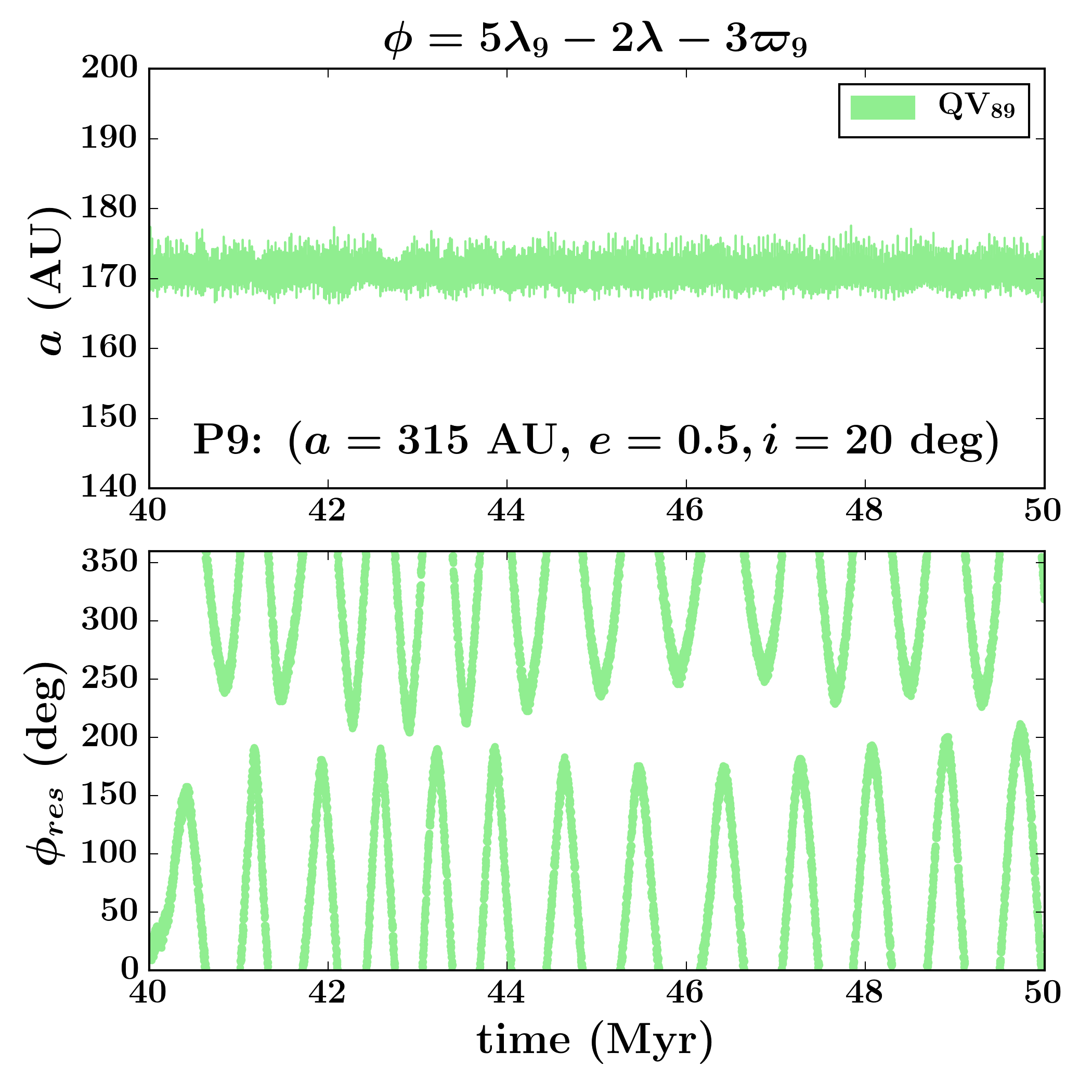}
\caption{Possible $5$:$2$ resonance between \ThingOne\ and Planet Nine. The top panel shows the time evolution of the semi-major axis for a clone of \ThingOne\ that lands in a $5$:$2$ resonance with Planet Nine. The bottom panel shows the corresponding resonance angle given by the argument $\phi = 5 \lambda_9 - 2 \lambda - 3 \varpi_9$. The resonance angle is clearly librating and thus indicates a mean motion resonance with Planet Nine.} 
\label{fig:resonance52}
\end{center}
\end{figure}

As one working example, Figure \ref{fig:resonance52} shows the behavior of a clone of \ThingOne\ over a time interval for which it resides in a $5$:$2$ resonance with Planet Nine. The top panel shows the nearly constant semi-major axis evolution that is characteristic of resonance, and the bottom panel shows the librating resonance argument $\phi$. In this plot, we only show the time interval during which this object is truly in resonance. However, \ThingOne\ often resides in near-resonance for billions of years, most likely experiencing nodding behavior \citep{2013ApJ...762...71K} in which it transitions back and forth from a librating to a circulating resonance argument (note that similar behavior has been reported in \citealt{sarah} and \citealt{Becker} for some of the longer-period ETNOs, such as Sedna). 

\begin{figure}[t!]
\epsscale{1}
  \begin{center}
      \leavevmode
\includegraphics[width=85mm]{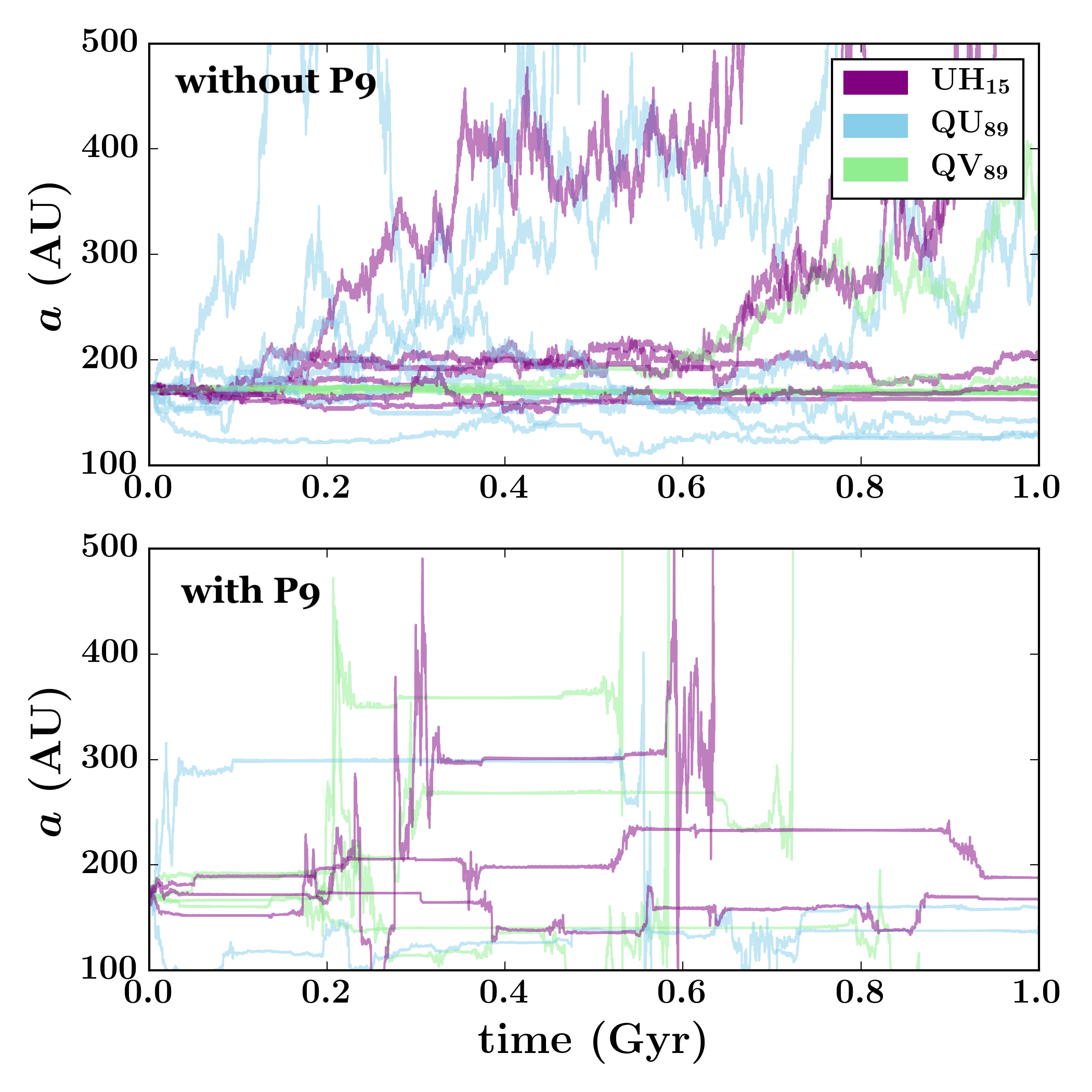}
\caption{The top panel shows the rapid semi-major axis evolution of a few representative clones of the triplet objects in the current Solar System. The bottom panel shows the constant-$a$ evolution induced by the addition of Planet Nine with $a_9 = 315$ AU, $e_9 = 0.5, i_9 = 20^\circ$ into the system.} 
\label{fig:a_comparison}
\end{center}
\end{figure}

An example of resonance hopping behavior is shown in the bottom panel of Figure \ref{fig:a_comparison}, which displays the evolution of the triplets in the presence of Planet Nine. It is clear that in this dynamical regime, these objects spend more time with constant semi-major axis than in the transient, highly variable regimes in between. In other words, the triplet objects effectively ``hop'' from one value of semi-major axis to another. 

In contrast, the top panel shows the evolution of the triplet objects in the presence of the current Solar System only (without Planet Nine). Here we see that the objects tend to spend more of their time scattering, with more highly variable semi-major axes, and only spend short time intervals with nearly constant orbital elements. In other words, rather than continually hopping between semi-major axis values, the objects only occasionally get ``stuck''  into a resonance. As such, this plot highlights the difference between the effects of ``resonance sticking" (top) and ``resonance hopping" (bottom)\footnote{For a recent numerical characterization of resonance sticking in the Kuiper Belt, see the analysis by \citet{2018arXiv180508228Y}.}. 

In light of this dynamical behavior, perhaps it is not as surprising to find a collection of ETNOs with similar semi-major axes. Rather than invoking collisions to explain the origin of this associated ETNO triplet, an alternate explanation is that these three objects are currently in the same mean-motion resonance with Planet Nine. Since resonances result in semi-major axis oscillations with a finite width, it is not necessary for the three ETNOs to have identical values of $a$ as long as they are similar. In fact, the observed differences of a few percent correspond to a relatively narrow libration amplitude of the resonance argument. 

Of course, we cannot know whether these objects are in fact in resonance with Planet Nine until the planet is discovered (or ruled out). For a given orbit of Planet Nine, however, we can search for librating resonance angles among the clones in our simulations. It is important to note that to achieve anti-aligned behavior for these  triplet objects, we have chosen Planet Nine parameters that fall on the inside of the typically accepted range of $(a_9, e_9)$. The original authors --- as well as subsequent analyses --- have shown that the perihelion distance of Planet Nine is likely to be $q_9 \sim 250$ AU, which is significantly larger than our choice of $q_9 \sim 150$ AU. Taking this difference into account, the next section considers a more typical set of parameters for Planet Nine. As a consequence, the triplet objects lose their anti-alignment in $\Delta \varpi$ space. 

\subsubsection{ETNOs with Circulating $\Delta \varpi$}
\label{sec:circres}

Given the current arguments $\varpi$ of the three objects under consideration, it is possible that their orbits are in fact circulating in $\Delta \varpi$ and do not remain confined in physical space. Given their observed semi-major axes, which are smaller than the typical values for which Planet Nine dominates TNO dynamics, the $\Delta{\varpi}$ for these ETNO orbits are likely to circulate even in the presence of Planet Nine. In this case, the currently observed anti-alignment with the proposed Planet Nine orbit is due to chance. Even with circulating $\Delta{\varpi}$, the dynamical behavior of the ETNOs could be driven by (i) resonant or (ii) secular interactions with Planet Nine, as discussed below. 

\begin{figure}[t!]
\epsscale{1}
  \begin{center}
      \leavevmode
\includegraphics[width=85mm]{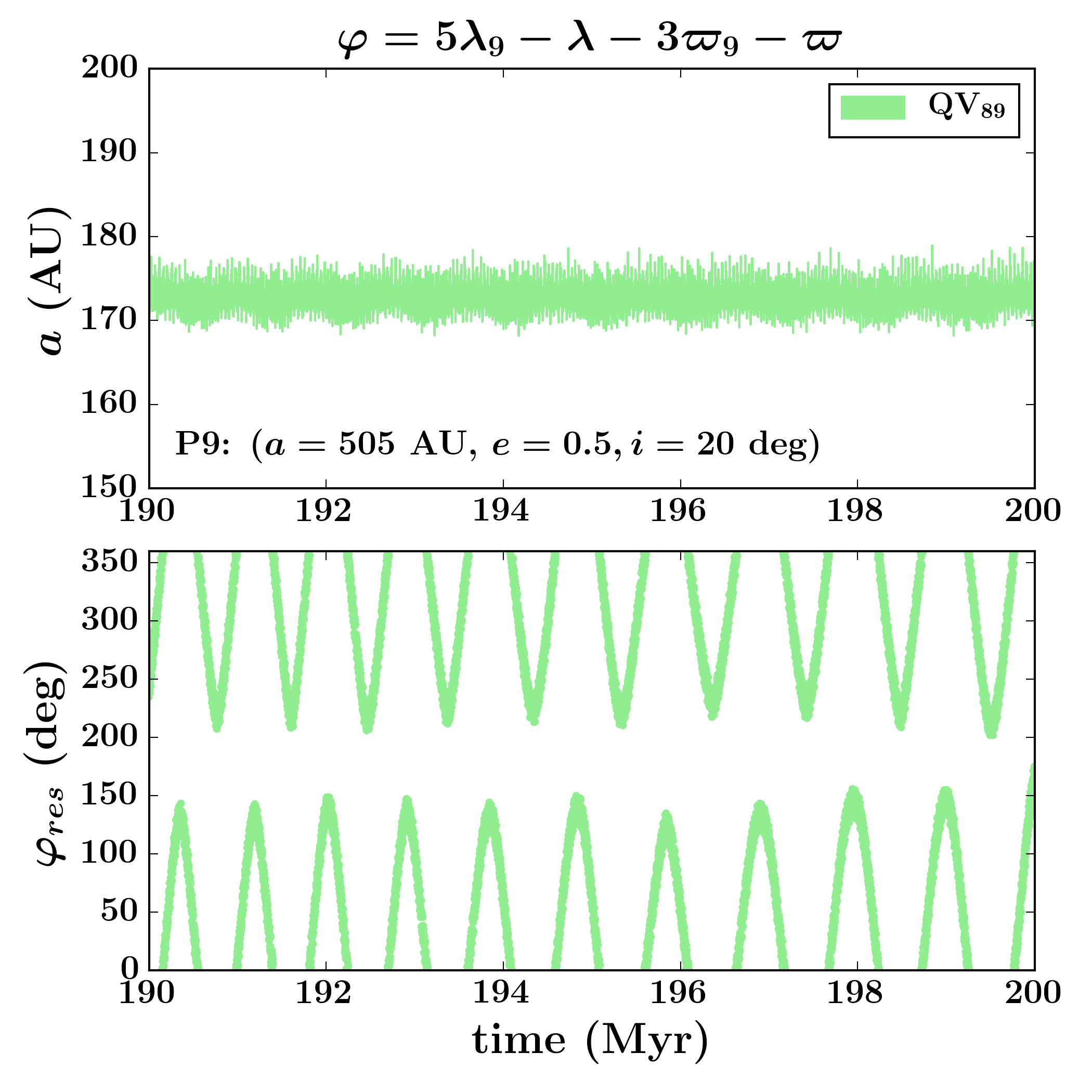}
\caption{Possible $5$:$1$ resonance between \ThingOne\ and Planet Nine. The top panel shows the time evolution of the semi-major axis for one clone of \ThingOne\ that lands in a $5$:$1$ resonance with Planet Nine. The bottom panel shows the corresponding resonance angle given by the argument $\varphi = 5 \lambda_9 - \lambda - 3 \varpi_9 - \varpi$. The resonance angle is clearly librating and thus indicates a mean motion resonance with Planet Nine. } 
\label{fig:resonance51}
\end{center}
\end{figure}

\textbf{Resonant Case.}
In theory, some ETNOs with circulating $\Delta \varpi$ could be trapped in mean-motion resonances with Planet Nine. Despite their potentially orbit-crossing behavior, these objects would be able to remain stable in the presence of Planet Nine due to the associated phase-space protection mechanism. In this case, the form of the librating resonance angle is different from the anti-aligned case \citep{BatMorb}. An example of an object with circulating $\Delta \varpi$ but which also has a librating resonance argument is shown in Figure \ref{fig:resonance51}.

If the triplet objects are in this dynamical class, then the similarity of their orbits could be explained by their mean-motion resonance with Planet Nine, as in section \ref{sec:antialigned} above for the anti-aligned class. 

\textbf{Non-Resonant Case.} Although objects circulating in $\Delta \varpi$ may be in mean-motion resonances with Planet Nine, not all must be in this synchronized state in order to avoid close encounters with Planet Nine. Depending on the exact orbit of the planet, it is possible for the objects with circulating $\Delta \varpi$ values to follow trajectories that avoid orbit-crossing configurations. In this case, the behavior of these objects is driven by secular interactions with Planet Nine, and although these objects are not in true resonance, their dynamics are marked by regions of constant semi-major axis evolution.

\section{Conclusion} 
\label{sec:conclude} 

This paper has reported the discovery of two new extreme trans-Neptunian objects, \ThingOne\ and \ThingTwo.
Our main findings can be summarized as follows:

[1] These two objects may be associated with each other and with the ETNO 2013 UH$_{15}$. The three objects have orbital elements $(a,e)$ that differ by less than about 2 percent among the three bodies, and the remaining angular orbital elements $(i,\omega,\Omega)$ fall within ranges of 9, 20, and $\sim70$ degrees (see Table \ref{table:1}). For \ThingOne\ and \ThingTwo, the observed colors are statistically indistinguishable
(see Table \ref{table:1}). However, we cannot state with certainty that the apparent simularities in orbits are not due to random chance. 

[2] The existence of this triplet ETNO system has important ramifications for the dynamics of the outer Solar System. Numerical integrations indicate that the three bodies tend to stay together in parameter space for time scales of order of 100 Myr, but diverge over longer spans of time (see Figure \ref{fig:aeievolution}). In addition, the probability that three random ETNOs happen to lie so close in orbital element space is low (see Figure \ref{fig:MPC_real_comp}), but this possibility cannot be excluded with current data. These objects should be considered a candidate system for a common origin in future analyses. 

[3] These objects are not likely to be in true resonances with Neptune at the current epoch. However, it is possible that they were in resonances in the recent past; this possibility could explain the present-day similarity in their semi-major axes. These objects appear to exhibit ``resonance sticking" in our simulations, generally residing outside of Neptune resonances but ``sticking" to resonances for short periods of time. 

[4] The existence of this triplet ETNO system has important implications for the Planet Nine hypothesis. The orbits of the three ETNOs under study should be sufficiently distant from the Sun that they would be influenced by the proposed Planet Nine, even though their orbital distances fall below the $a \sim 250$ AU cutoff used previously in some works \citep[ex:][]{sarah, Becker, BatMorb} to describe the population most susceptible to Planet Nine's influence. In particular, in the presence of Planet Nine, with canonical orbital elements and mass, the long-term evolution of semi-major axes for the three bodies is markedly different, as they exhibit ``resonance hopping" whereby long stints in resonance are disrupted by short bursts of migration in semi-major axis. In addition, if Planet Nine exists with a semi-major axis near the low end of its proposed range, then it can cause the ETNO orbits to be apsidally anti-aligned (as observed). 
The apparent alignment of the physical orbits of these objects with that predicted by Planet Nine requires future work to determine the true extent of Planet Nine's dynamical influence, as the 250 AU cutoff may not be sufficient to describe the population of objects shepherded by Planet Nine. 


Future discoveries and their associated dynamical studies will expand the census of known ETNOs and allow for a better determination between the various scenarios presented in this work. 

\vskip 0.5truecm

This material is based upon work supported by the National Aeronautics and Space Administration under Grant No. NNX17AF21G issued through the SSO Planetary Astronomy Program, and by NSF grant AST-1515015. We thank Konstantin Batygin, Mike Brown, Sarah Millholland, Andrew Vanderburg, John Monnier, and Kat Volk for useful conversations. We thank the anonymous referee for valuable comments that increased the clarity of the manuscript. JCB and SJH are also supported by the NSF Graduate Research Fellowship Grant No. DGE 1256260. The computations for this work used the Extreme Science and Engineering Discovery Environment (XSEDE), which is supported by National Science Foundation grant number ACI-1053575. This research was done using resources provided by the Open Science Grid, which is supported by the National Science Foundation and the U.S. Department of Energy Office of Science.

Funding for the DES Projects has been provided by the U.S. Department of Energy, the U.S. National Science Foundation, the Ministry of Science and Education of Spain, 
the Science and Technology Facilities Council of the United Kingdom, the Higher Education Funding Council for England, the National Center for Supercomputing 
Applications at the University of Illinois at Urbana-Champaign, the Kavli Institute of Cosmological Physics at the University of Chicago, 
the Center for Cosmology and Astro-Particle Physics at the Ohio State University,
the Mitchell Institute for Fundamental Physics and Astronomy at Texas A\&M University, Financiadora de Estudos e Projetos, 
Funda{\c c}{\~a}o Carlos Chagas Filho de Amparo {\`a} Pesquisa do Estado do Rio de Janeiro, Conselho Nacional de Desenvolvimento Cient{\'i}fico e Tecnol{\'o}gico and 
the Minist{\'e}rio da Ci{\^e}ncia, Tecnologia e Inova{\c c}{\~a}o, the Deutsche Forschungsgemeinschaft and the Collaborating Institutions in the Dark Energy Survey. 

The Collaborating Institutions are Argonne National Laboratory, the University of California at Santa Cruz, the University of Cambridge, Centro de Investigaciones Energ{\'e}ticas, 
Medioambientales y Tecnol{\'o}gicas-Madrid, the University of Chicago, University College London, the DES-Brazil Consortium, the University of Edinburgh, 
the Eidgen{\"o}ssische Technische Hochschule (ETH) Z{\"u}rich, 
Fermi National Accelerator Laboratory, the University of Illinois at Urbana-Champaign, the Institut de Ci{\`e}ncies de l'Espai (IEEC/CSIC), 
the Institut de F{\'i}sica d'Altes Energies, Lawrence Berkeley National Laboratory, the Ludwig-Maximilians Universit{\"a}t M{\"u}nchen and the associated Excellence Cluster Universe, 
the University of Michigan, the National Optical Astronomy Observatory, the University of Nottingham, The Ohio State University, the University of Pennsylvania, the University of Portsmouth, 
SLAC National Accelerator Laboratory, Stanford University, the University of Sussex, Texas A\&M University, and the OzDES Membership Consortium.

Based in part on observations at Cerro Tololo Inter-American Observatory, National Optical Astronomy Observatory, which is operated by the Association of 
Universities for Research in Astronomy (AURA) under a cooperative agreement with the National Science Foundation.

The DES data management system is supported by the National Science Foundation under Grant Numbers AST-1138766 and AST-1536171.
The DES participants from Spanish institutions are partially supported by MINECO under grants AYA2015-71825, ESP2015-66861, FPA2015-68048, SEV-2016-0588, SEV-2016-0597, and MDM-2015-0509, 
some of which include ERDF funds from the European Union. IFAE is partially funded by the CERCA program of the Generalitat de Catalunya.
Research leading to these results has received funding from the European Research
Council under the European Union's Seventh Framework Program (FP7/2007-2013) including ERC grant agreements 240672, 291329, and 306478.
We  acknowledge support from the Australian Research Council Centre of Excellence for All-sky Astrophysics (CAASTRO), through project number CE110001020, and the Brazilian Instituto Nacional de Ci\^encia
e Tecnologia (INCT) e-Universe (CNPq grant 465376/2014-2).

This manuscript has been authored by Fermi Research Alliance, LLC under Contract No. DE-AC02-07CH11359 with the U.S. Department of Energy, Office of Science, Office of High Energy Physics. The United States Government retains and the publisher, by accepting the article for publication, acknowledges that the United States Government retains a non-exclusive, paid-up, irrevocable, world-wide license to publish or reproduce the published form of this manuscript, or allow others to do so, for United States Government purposes.

\software{WCSfit (Bernstein et al. 2017), mp\_ephem  (https://github.com/OSSOS/liborbfit), pandas \citep{ mckinney-proc-scipy-2010}, IPython \citep{PER-GRA:2007}, matplotlib \citep{Hunter:2007}, scipy \citep{scipy}, numpy \citep{oliphant-2006-guide}, Jupyter \citep{Kluyver:2016aa}}

\bibliographystyle{apj}

\end{document}